# A Critical Review of Traffic Signal Control and A Novel Unified View of Reinforcement Learning and Model Predictive Control Approaches for Adaptive Traffic Signal Control

Xiaoyu Wang, Baher Abdulhai
Department of Civil Engineering
University of Toronto
Toronto, Canada
cnxiaoyu.wang@mail.utoronto.ca, baher.abdulhai@utoronto.ca

Scott Sanner
Department of Mechanical and Industrial Engineering
University of Toronto
Toronto, Canada
ssanner@mie.utoronto.ca

*Abstract*—Recent years have witnessed substantial growth in adaptive traffic signal control (ATSC) methodologies that improve transportation network efficiency, especially in branches leveraging artificial intelligence based optimization and control algorithms such as reinforcement learning as well as conventional model predictive control. However, lack of cross-domain analysis and comparison of the effectiveness of applied methods in ATSC research limits our understanding of existing challenges and research directions. This chapter proposes a novel unified view of modern ATSCs to identify common ground as well as differences and shortcomings of existing methodologies with the ultimate goal to facilitate cross-fertilization and advance the state-of-the-art. The unified view applies the mathematical language of the Markov decision process, describes the process of controller design from both the world (problem) and solution modeling perspectives. The unified view also analyses systematic issues commonly ignored in existing studies and suggests future potential directions to resolve these issues.

*Keywords—Adaptive traffic signal control, dynamical systems, Markov decision process, artificial intelligence, reinforcement learning, model predictive control*

## I. Introduction

Transportation systems play an important role in society by providing mobility and accessibility for people and goods. Increasing populations and urbanization pose escalating burden on existing traffic networks and exacerbate traffic congestion, especially in large cities and metropolises such as the Great Toronto and Hamilton Area (GTHA). The negative impacts of congestion are manifold (SAUNDERS 2014), including but not limited to productivity loss, fuel and energy wastage, increased pollution, and safety risks. Metrolinx, an agency of the Government of Ontario, estimates a $6 billion annual costs caused by traffic congestion in the GTHA in 2008 (Metrolinx, 2008). (Dachis 2013) develops a new approach to evaluating infrastructure investment and adjust the estimated annual cost to $7.5 to $11 billion.

Alleviating traffic congestion requires efforts from multiple aspects: regional transportation planning, public transit expansion and integration, and road construction. However, the realization of any one of them involves a long and expensive process of infrastructure upgrades, which may be limited by budget and space constraints as well as environmental and sustainability concerns. How to maximize transportation efficiency given the current road network resources naturally becomes an important alternative to brute force infrastructure expansion. Intelligent transportation systems (ITS) aim to provide high-quality services and make the network coordinated and smarter with the help of emerging sensing, communication, information, and control technologies. Among the key ITS functions, adaptive traffic signal control (ATSC) aims to maximize the use of current road capacity by adjusting traffic signal timing plans in real-time according to the sensed traffic dynamics and grant intersections the ability to respond to demand variations to reduce intersection delays and queues and maximize throughput. Moreover, traffic signal coordination targets network-level optimality via coordinating decision making among a number of intersections in one-dimension such as on a corridor or in two-dimensions such as a grid network.

In order to achieve this goal, decades of research derived many classical solutions serving traffic systems all around the world till today (such as SCATS (Sims and Dobinson 1979) and SCOOT (Hunt et al. 1981)), and led to pioneer studies exploiting the potential of adopting modern optimal control and reinforcement learning (RL) achievements. The newer methods that lean on emerging techniques claim better performance than older approaches, but lack cross-evaluation and in-field tests. Among them, RL-based methods draw their appeal on the basis that they do not require calibrated system models, i.e. they are model-free approaches that can learn directly from interacting with simulated environment. Compared with classical ATSCs, which require efforts on problem modeling in either explicit (e.g. model predictive control, MPCs) or implicit (e.g. approximated model learned by neural networks) manner, RL-based methods seem easier to use. Fig. 1.1 illustrates their differences, where in MPC the problem modeling stage refers to building a high fidelity model that might be too complex to establish and/or apply, and the model simplifying stage refers to the optional process of simplifying the former stage's output till a feasible level of complexity. The common routine of RL-based methods replaces the modeling stages in the MPC-based approaches with the state space designing stage and conduct this stage with feature engineering. The succinctness of the RL approach shown in the right-hand side routine in Fig. 1.1 has drawn significant appeal in recent years. However, the performance of RL methods ties closely to satisfying key assumption: the Markov decision process model described by the solution must satisfies the requirements of system stationarity and observability. For real world problems with complex dynamics, such as in ATSCs, a solution that breaks the forementioned assumptions may in turn lose the performance guarantees that the RL theory provides. Furthermore, differences in language between the MPC and RL disciplines make cross-comparison of recently developed methods a formidable challenge.

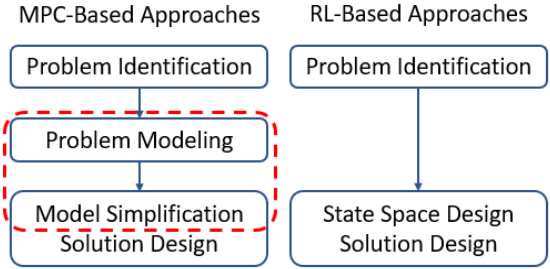

Figure 1.1. Typical procedures for designing MPC and RL-based solutions. The red dashed box identifies their differences.

In ATSC, the above issues can be summarized into gaps among methodologies (e.g. MPC vs RL), and gaps between theory and reality (e.g. satisfying the Markovian assumptions). Hence, before conducting breakthrough research, it is crucial to fill these gaps, identify the challenges, and clarify the research directions. For this purpose, this article proposes the unified view describing generic traffic signal control (TSC) problems in the form of Mixed Observability Markov Decision Processes (MOMDP), discusses TSC solutions with such a unified language, and identifies a viable direction for RL designing to address the gaps. Specifically, from an RL's perspective, Section III answers the problem modeling question, and Section IV fills up the missing model simplification procedure in RL as shown in Fig. 1.1. In other words, the proposed unified view identifies current challenges in existing model-free reinforcement learning (MF-RL) methods for TSC, involving but not limited to the issues of partial observability and non-stationarity from exogenous disturbance and multi-agent policies.

The unified view suggests directions for mitigating the above issues by agent modeling and identifies the potential difficulties in deploying CTDE (centralized training, decentralized executing) methods. Further, the fixed policy derived from RLs commonly cannot describe the full decision (state-action) space owing to the limited visitable area during training restricted by the experience. The experience, described with probabilistic distribution meaning the experienced system dynamics, is largely determined by the demand pattern and behavior policies (the agent's control strategies). Once the learned policy gets deployed in the field, it will be required to make decisions within the unforeseen part of the decision space, in which the policy is not well-tuned to provide a reasonable decision. That depicts the distribution shifting issue in RL: a distributional shift between training and testing data. The unified view suggests alleviating such an issue by enhancing the learned policy with online planning, which gives more reliable decisions to the under-learned scenarios (also referring to the generalizability in AI). Finally, hierarchical control is suggested for improving coordination in long-horizon large-scale network scenarios.

The remainder of this article is organized as follows: Section II gives problem background and literature review, Section III introduces necessary preliminaries and builds up the unified view for TSC problems, Section IV critiques existing methodologies under the proposed unified view and leads to the research targets, and Section V presents potential research directions guided by the unified view.

## II. BACKGROUND AND LITERATURE REVIEW

In this section, we formulate the traffic signal control (TSC) and coordination problem and present a generic categorization of potential solutions. Then, we provide an overview of the history of TSC methodologies and provide a critical review of types of solutions, which leads to the objectives of this article.

## A. Problem Formulation

### 1) Traffic Signal Control: Network Scale, State, Action and Performance

Given a group of to-be-controlled intersections, which may form a single or multiple regions (sub-networks), within a traffic network, we aim to optimize the network efficiency identified by performance criteria through adjusting intersections' traffic signal timing plans. Despite some literature studying the single intersection control problem specifically, we treat it as an extreme case of the more general multi-intersection control problem because they largely share the common problem definition and can be modeled in the same manner.

We further discuss the TSC problem from three perspectives of a system: state perception, action execution, and performance evaluation.

**State perception**

The optimized traffic signal timing plans can be either fixed-time or time-varying adapting to traffic conditions (adaptive for short). Fixed time plans are determined using historical traffic averages over a given period of the day and repeat throughout that period of the day. The adaptive plans can be determined either prior to the deployment based on time-dependent historical data (open loop) or in real-time based on live traffic measurements (closed loop). Besides their methodology differences covered by Section II.B, another key difference between the two is the requirement on the sensing ability of the traffic status. Fixed-time controllers only rely on the network topology and the historical observation of traffic such upstream flows, stop-bar turning counts, link densities and speeds etc. Most of these requirements are easy to be fulfilled without the expenses of deploying complex sensing infrastructures (Gordon et al. 1996).

To the contrary, real-time controllers rely on not only much elaborate real time sensing but also communicating and computing capabilities, which may be a restriction to the deployment of some modern TSC solutions. Limited by the scope and length of this article, we briefly introduce some current and promising future sensing technologies. A list of potential elements to form a state space can be found in Appendix I.

A commonly used sensor is the inductive loop detector (Gordon et al. 2005) deployed under the road pavement which produces signal change when vehicles are above it. It provides a reliable observation at both the microscopic scale (detector occupancy and vehicle speed) and the macroscopic scale (average traffic flow and average speed). But, restricted by its fixed position at a given point on the road, the information it could collect is limited. With the development of computer vision (CV) technology and the broader deployment of traffic surveillance cameras, we can expect to get access to more traffic information in real-time: identifying all vehicles' positions and speeds within cameras' detection range with image processing algorithms (Tang et al. 2019). What's more, vehicle-to-vehicle (V2V) and vehicle-to-infrastructure (V2I) communication technologies are also under rapid development. With V2V/V2I communication, vehicles share their exact states (position, speed, etc. provided by GPS, odometer, and inertial navigation systems) to the roadside units (RSU) or even the traffic information center (Dey et al. 2016) from anywhere within the traffic network. Other detection technologies are also emerging such as long-range radar and lidar (light detection and ranging) but they are less common in practice so far.

**Action execution**

An intersection consists of a collection of (conflicting) movements — connections from upstream incoming lanes to downstream outgoing lanes. Granting right-of-way (ROW) and green lights to a group of movements forms a phase. All phases together with intermediate yellow and all-red times form a complete cycle, that is, a timing plan. A group of phases with fixed movement combinations forms a phasing scheme. Optimizing the timing plan involves the optimization of the phasing scheme (Gordon et al. 1996, 3–15 — 3-19), (Koonce and Rodegerdts 2008). However, considering the complexity of optimizing all elements within a timing plan, in this article, we assume a given phasing scheme and optimize the order and duration of all candidate phases only.

We categorize the timing plan optimization problem given a fixed phasing scheme into two folds: second-based and cycle-based, namely, taking actions every second(s) or every cycle(s) respectively.

Second-based controllers make a decision every one or a few seconds, i.e. the time step is one or a few seconds. Every time step, the controller selects the next phase from the candidate list of phases. The controller maintains the current phase if the same phase is re-selected and meets the phase duration constraints; otherwise changes to the selected phase and inserts yellow and all-red change intervals. Second-based controllers can be further divided into fixed phasing sequence (FPS) and variable phasing sequence (VPS) (El-Tantawy 2012, 109). FPS candidate set contains only the current phase and the next phase in an order that is given by the phasing scheme. The decision is therefore binary: "extend" the current phase or "change" to the next phase. VPS candidate set contains all possible phases. The VPS action space provides more flexibility to the controller and leads to better performance in simulated scenarios (El-Tantawy 2012, 140–41). Although second-based VPS can be most efficient and most flexible, the VPS aspect may be confusing to and unfavored by some drivers as the next selected phase may not conform to their expectations. In addition, in second-based control drivers cannot be informed of when the next phase is coming or how much time is left before they receive green.

Cycle-based controllers determine the timing plan for cycles directly. Given a phasing scheme, a controller can adjust one or more of three main elements for the next one or a few cycles: the phase order, the splitting representing the proportion of green time allocated to each phase, and the cycle length, as shown in Fig 1.2. This type of controller is less flexible than the second-based controller but more popular in the industry since drivers know exactly the length of each phase and can anticipate their waiting time more accurately. Besides, cycle-based controllers are more pedestrian-friendly. Taking pedestrian lights in Canada as an example, a determined phase length could provide maximum "Walk" signal length; while second-based controllers extending each phase one second at a time provide a short "Walk" signal followed by a "Flashing Don't Walk" (Koonce and Rodegerdts 2008) as the phase can terminate at any second after a minim green period.

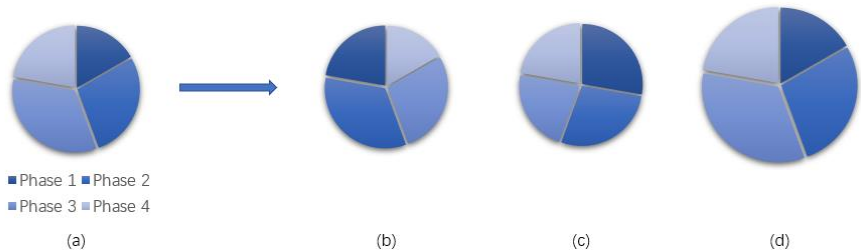

Figure 1.2. Traffic light cycles represented with pie charts. (a) A four-phase cycle. (b) Changed phase order. (c) Changed split. (d) Changed cycle length.

**Performance evaluation**

We evaluate the traffic network efficiency from multiple perspectives: intersection-level, network-level, and arterial-level. We briefly discuss the usage of each category and leave formal definitions to Appendix II.

Intersection-level criteria are commonly used in reward / loss functions for the TSC problem and include, for example, intersection delay, queue lengths etc. Estimating a criterion in real-time requires elaborate detection capability. Despite traffic managers wanting to optimize a network-level global target, it is intractable or over expensive to do so in most cases. Hence, it is more common to optimize and evaluate the performance of single intersections and coarsely estimate global network level targets by summing up local criteria over multiple intersections.

Optimizing a network globally, on the other hand requires network-wide sensing and optimization and the control problem itself grows intractable with the number of intersections in the network being optimized. Network-level criteria, such as total travel time in the network, can be more easily measured in simulations rather than in the field, hence are commonly used as measurements of effectiveness (MOEs) for evaluation.

Different from the above two types, arterial-level criteria are often defined to describe the quality of signal progression (Manual 2000) among two or more consecutive intersections to allow vehicles to pass multiple intersections without stopping. The wider the green band and the larger the number of intersections that vehicles progress through the better. But, due to the "closure" effect of urban traffic networks (Roess, Prassas, and McShane 2004), signal progression indicators in grid networks are hard to be optimized simultaneously. Hence, this type of criteria is widely used in heuristic methods in the traffic field but can be rarely found in modern optimization-based solutions. The formal definition of signal coordination (progression) and its disambiguation are discussed next in Section II.A.2.

*2) Traffic Signal Coordination*

We believe there is confusion in the usage of the term "coordination" in the TSC field. In this section we offer clarifying definitions of the term to resolve the ambiguity and confusion.

**Intersection coordination, signal progression and green waves**

In the traditional traffic literature, intersection coordination, also known as signal progression, refers to the problem of coordinating the onset of green times at successive closely spaced intersections to move vehicles efficiently through the set of signals with minimal stopping. It is a clear objective of the generic signal timing optimization problem (see Appendix II.C) and has been broadly used in classic offline optimized traffic control solutions (Little, Kelson, and Gartner 1981; Gartner et al. 1990). For disambiguation purposes, we term it **progression**.

**Agent (controller) coordination**

A broader definition of "coordination" can be found in more generic fields including planning, control, game theory (where mostly called player cooperation), etc. Coordination in this context refers to the problem of finding the global optimum under a certain set of objectives by leveraging communication and decision-making coordination among individual local controllers or agents. It is easy to conclude that signal progression, as an objective, is a subset of agent coordination. For disambiguation purposes, we use the term **coordination** for this broader concept of agent-coordination. Types of control architectures for reaching this coordination will be discussed in the next section.

*3) Paradigms of Control Architectures*

The complexity of finding the global optimum timing plans for a group of intersections in a network grows exponentially with the number of intersections involved. This implies the necessity of problem decomposition. According to the structures of control architectures, we categorize network-wide signal control into three paradigms: centralized, decentralized and hierarchical. An overview comparison diagram can be found in Fig 1.3, where each controlled region could represent single or multiple intersections.

**Centralized control**

As shown in Fig 1.3(a), in a centralized control system, a single controller governing all target intersections receives the joint system state and makes the joint decision (the action vector corresponding to all controlled objects). This type of approach models the entire system including the potential interaction among intersections, which provides the opportunity of finding the global optimum. However, the dimensionality of the problem renders the optimization of the control too complex or intractable and limits the probability of reaching the global optimum in an acceptable time. The centralization also leads to high communication costs and low fault tolerance. As a consequence, the centralized paradigm can be merely found in online optimization approaches (Negenborn and Maestre 2014). Despite some decomposition techniques can be adopted to alleviate the complexity problem, such as factored spaces (Raghavan et al. 2012) transforming a joint decision making problem into a sequential decision making problem, most of these types of approaches still have to seek a balance between the model complexity and accuracy and is difficult to scale up.

**Decentralized control**

Fig 1.3(b) presents the decentralized control paradigm in which a group of controllers is responsible for their respective regions. Its divide and conquer idea alleviates the complexity issue that centralized controllers face. But it relies heavily on inter-controller communication (agent coordination), because local models cannot describe the full system dynamics and cannot capture the mutual influence among local controllers as good as a global model.

**Hierarchical control**

While decentralized control encounters the dilemma of agent communication and coordination, hierarchical control gives another option for finding the global optimum. As shown in Fig 1.3(c), an upper-level controller takes charge of a set of lower-level controllers and guides their decision-making in multiple ways: feeding future predictions (Head, Mirchandani, and Sheppard 1992), posing action constraints (Head, Mirchandani, and Sheppard 1992; Sims and Dobinson 1979), and providing optimization targets (Christen et al. 2021). Lower-level controllers in this paradigm can still communicate with each other for acquiring better knowledge for making actual decisions; while the upper-level controller runs on a large time-scale and focuses on longer-term planning.

Hierarchical control can have multiple layers depending on the way of decoupling the problem. It is worth noting that some centralized control systems with decomposition techniques may have similar layered structures (Kong et al. 2011), especially in the deep reinforcement learning field (Tan et al. 2019), but do not follow the hierarchical control paradigm — there are no upper-level controllers guiding lower-level ones.

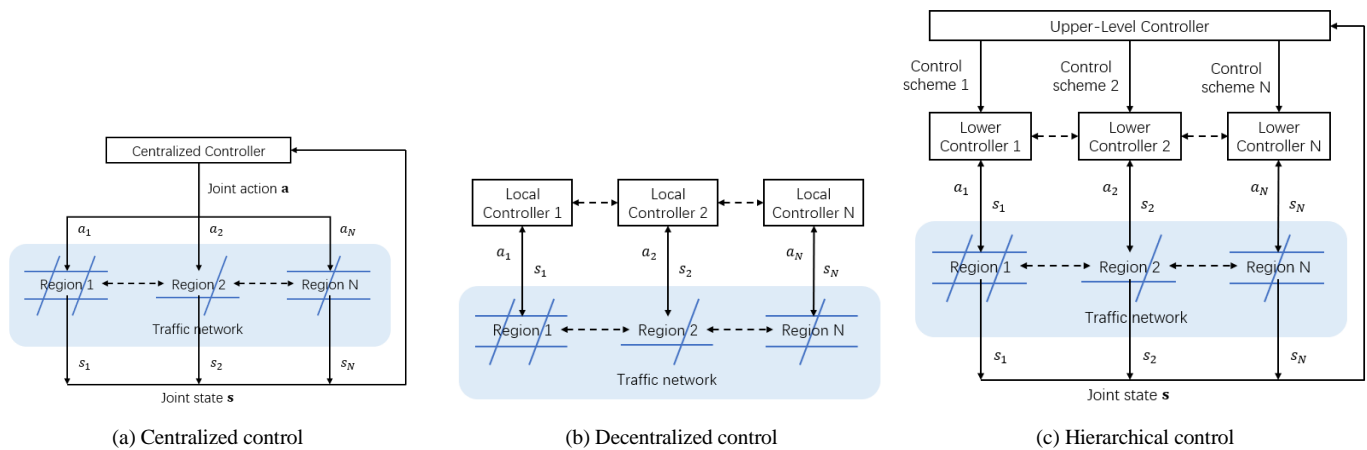

(a) Centralized control  (b) Decentralized control  (c) Hierarchical control

Figure 1.3. Three basic paradigms of control architectures, where dash arrows represent potential interactions among neighboring regions or controllers, and blue lines demonstrate example subnetworks that can have different topologies

*B. Literature Review*

As the early TSC research can be dated back to a half-century ago (e.g. Webster 1958), the TSC literature is massive. Over time, different strategies emerged and different methodologies have been discussed and implemented. In this section, following the historical trajectory of TSC development, we give a brief review with a major focus on recent progress.

*1) Overview of Historical Developments*

TSC solutions can be categorized into three domains: fixed-time control, actuated control, and adaptive control.

**Fixed-time control**

The optimized traffic signal timing plans can be either fixed-time or time-varying adapting to traffic conditions (adaptive for short). Fixed time plans are determined using historical traffic averages over a given period of the day and repeat throughout that period of the day. One of the most well-known algorithms in this area is the Webster method (Webster 1958) which assigns green time to each approach proportionally to serve corresponding traffic demand and minimize delay.

The Webster method is a fully decentralized approach without considering the interaction with and influence from neighbors. Alternatively, fixed-time control can also deal with regional problems, usually using a centralized approach. We classify them into two categories according to their optimization objectives: progression-based control and delay-based.

1. Progression-based methods aim to maximize the width of a green band (see Appendix II.C) along an artery. Intersections share the same cycle length but can have different splits and achieve desirable signal progression by adjusting their offsets — given a base intersection, the offset of another intersection means its cycle delay relative to the base intersection. The following are progression-based examples:

   - MAXBAND (Little, Kelson, and Gartner 1981) calculates cycles using the Webster method and optimizes bi-direction bandwidths with mixed-integer linear programming (MILP).
   
   - MULTIBAND (Gartner et al. 1990, 1991) extends MAXBAND and introduces multiple objectives besides the volume-weighted progression: reducing delay and stops. It also uses a MILP approach and provides more decision capability to the traffic system managers.
   
   - Progression analysis and signal system evaluation routine (PASSER) developed by the Texas Transportation Institute (Venglar, Koonce, and Urbanik II 2000).

2. Delay-based methods target a different objective and are not limited to artery scenarios. These approaches also share a common cycle length across considered intersections, and optimize their splits and offsets according to the built-in macroscopic delay estimation models. Examples are:

   - TRAffic Network StudY Tool and its extensions (TRANSIT, TRANSIT-7F, etc.) (Robertson 1969) optimizing arterial progressions.
   
   - SYNCHRO (Trafficware, n.d.) regional control functionality. SYNCHRO also supports offset and progression optimization.

Fixed-time control relies on information such as road network layout including the intersection configuration, historical traffic demand, traffic models, etc. Despite the network layout remains unchanged after deploying the timing plan, traffic dynamics vary considerably microscopically within cycles (micro dynamics), and macroscopically within day (peak vs. off-peak) and day to day and weekday to weekend. Hence, a major drawback of fixed time control is that it can't react to any demand fluctuation and dynamic variations. This fixed-time approach is suitable for intersections facing relatively stationary traffic demand. As an improvement, multiple fixed time plans can be developed for different periods of weekdays and weekend traffic.

**Actuated control**

Actuated control makes a signal plan more flexible and reacting to the fluctuation of traffic demand in real-time. In general, this type of method optimizes offline and adjusts phases heuristically online. We derive a control scheme including maximum green time for each phase based on historical information like in fixed-time control. While optimized green times are considered as maximums, the actual green time varies with sensor readings which indicate the presence of incoming vehicles from different approaches. If traffic ceases to arrive at the detector, the signal phase 'gaps out', but if traffic continues to arrive without gaps the green time is extended up to the predetermined maximum and the signal phase 'maxes out'. This type of controller is more flexibility than fixed-time control. But it is relatively myopic and focusing on the traffic arrivals on the active approach only without overall intersection optimization, hence has limited performance in complex scenarios (Roess, Prassas, and McShane 2004, 447–64). Actuated control, therefore, does not always lead to optimal performance.

**Adaptive control**

Adaptive traffic signal control (ATSC) emerged decades ago and introduced to dynamically optimize the traffic signal in light of the traffic dynamics, in finer time-scale, and continues under rapid development today. ATSC plans change over time to mirror traffic demand. The plan can be determined either prior to the deployment based on time-dependent historical data (open loop) or in real-time based on live traffic measurements (closed loop). ATSC methods always adjust the signal plan with time and traffic pattern changes, no matter the optimization process is done online or not. According to its development chronology, we categorize ATSC approaches into three stages: first generation, second generation, and modern optimized ATSCs. This section briefly introduces the first two stages and leaves modern techniques to the next sub-section.

1. First-generation (1G) ATSC methods are more like an extension of fixed-time control. They calculate multiple signal plans offline according to the traffic dynamics of different time periods of a day: morning peak, evening peak, and off-peaks. These controllers still cannot react to frequent and highly varying traffic demand fluctuations and need updating while long-horizon traffic dynamic varies. Some representatives are:

- Traffic responsive control (TR2) (Marchand et al. 1974).
- Urban traffic control systems (UTCS-1) (MacGowan and Fullerton 1979).

2. Second-generation (2G) ATSC methods solve the optimization problem online and provide more concrete reactions to the real-time traffic dynamics.

    - Sydney coordinated adaptive traffic system (SCATS) (Sims and Dobinson 1979), a hierarchical control system, was developed in the early 1970s in Australia. Its upper-level controller optimizes general control scheme such as cycle length and splits based on current traffic states; while the lower-level controllers, who are in charge of individual intersections, determines exact phases constrained by the upper-level posed control scheme, like an actuated controller. SCATS is not only delay-focused, but it also can achieve signal progression by the regional controller covering intersections in a corridor.
    - Split, cycle, and offset optimization technique (SCOOT) (Hunt et al. 1981) is another hierarchical control system invented in the 1970s and deployed by the British in 1980. At the higher-level, each SCOOT controller is responsible for a region, predicts the traffic profile with readings from upstream inductive loop detectors, and optimizes cycle lengths. Node-level controllers then adjust offsets and time splits locally. It can be seen as an online optimization version of the TRANSYT and SYNCHRO systems.

    The second-generation control systems are the most representative and widely used ATSC systems across the world: UK, USA, Canada, Australia, China, etc. Researchers carried out studies evaluating the performance and benefits these systems could provide (Bretherton 1990). Reports indicate its improvement on traffic delay compared to fixed-time systems, and also identifies its limitation under high density, highly dynamic traffic scenarios. Moreover, these systems still focus on historical and present information and do not try to consider the mutual influence between traffic dynamic and actions in the future.

Some previous literature may argue that second-generation systems' centralized and hierarchical paradigms are not reliable due to their communication cost and potential communication failure. The counter opinion is that, considering the development in related fields (as we discussed in Section II.A.1), communicating necessary local information to the system center is no longer intractable. Hence, the need for long-range communication and the centralized and hierarchical paradigms should not be ruled out.

*2) Modern Optimized Control: Emerging ATSC Techniques*

After the 1980s, many modern ATSC systems emerged, but few of them are deployed and tested in the field. On the other hand, despite the above classical approaches more or less have their drawbacks, they are widely deployed and are still playing important role around the globe. In this sub-section, we briefly introduce emerging methodologies in the literature and leave the theoretical analysis to Section III and IV.

Before continuing, we give an intuitive definition to the concept "policy", which will be formally defined in Section III.A. Given a system, a policy is a fixed scheme or a control law guiding the controller making decisions based on the sensed system state in either a deterministic or probabilistic manner.

**Rule-based heuristic policy**

A well known example of such methods is the max-pressure policy (Varaiya 2013; Anderson et al. 2018). It provides a decentralized policy aiming to enlarge the network throughput, rather than locally, and adjusts the signal plan online according to the real-time traffic dynamic. It is fully decentralized and tries to balance queue lengths within different intersections to prevent spill-back and over-saturated situations, which may further cause large-scale congestion in the network. (Anderson et al. 2018) shows its stability and gives its stability region — under a level of traffic load, the policy could stabilize the traffic network and maintain queues limited. It reacts to the fluctuation timely and faster than 1G and 2G ATSCs. But its policy is not optimized based on a given traffic model, which leaves spaces for pursuing the optimality.

Readers may find that the max-pressure policy is similar to other broadly used heuristic baselines in the TSC field such as the max-queue-first policy, a second-based controller that grants green light to the phase corresponding to the maximum queue length. The latter policy performs distributed strategy myopically in each intersection and does not consider coordination at all. The myopic property makes it fails under saturation scenarios, where the max-pressure policy does better.

**Computational intelligence techniques**

Computational intelligence (CI) refers to a collection of biologically and linguistically motivated computational methods. Traditional CI has some overlaps with other research fields and commonly includes fuzzy systems, evolutionary computation, and artificial neural networks (ANNs). Besides the well-known ANNs, fuzzy systems are inspired by the human language model imprecision in logic and solve problems with uncertainty. Fuzzy controls adopting fuzzy logic perform approximate reasoning relying on human knowledge and can be combined with ANNs. Evolutionary computation inspired by biologic evolution solves optimization problems heuristically without a requirement of the model. Examples such as genetic algorithms and swarm intelligence are widely adopted in some complex optimization tasks where model-based optimization is hard to deal with.

CI methods relax the dependency on explicit models and have been introduced to the ATSC field for decades. (Zhao, Dai, and Zhang 2011) overviews CI applications in both surface network and freeway network scenarios. The drawback of CI is clear, especially for fuzzy systems and evolutionary computation, that these heuristic-based methods are not guaranteed to find an optimum.

**Model-based rolling optimization: planning, scheduling, and model predictive control**

This type of method is commonly identified as the third-generation of ATSCs in the literature (Gartner, Stamatiadis, and Tarnoff 1995) to separate with some early work and model-free reinforcement learning methods. However, this branch is still under rapid development and contains much recent progress. We combine them in the same section and do not split the methodologies.

Rolling optimization (or rolling-horizon optimization, RHO) represents a large class of methodologies solving discrete-time sequential decision-making problems in dynamic systems. What distinguishes rolling optimization from the fore-mentioned optimization approaches is that the rolling optimization utilizes the system dynamic model in a different way: it projects the influence of the current decision into the future. Given a fixed prediction horizon (commonly tens of times longer than the decision time step), at each decision point, the optimizer predicts the evolution of system dynamics till the end of the horizon based on the given model, optimizes a target function involving the impact from this full horizon, picks and executes the immediate decision from the optimized action trajectory, rolls the horizon one step forward and repeats this procedure (Xi and Li 2019).

In urban areas, traffic dynamics may vary frequently due to the high volume and non-stationarity. Within a region, urban traffic networks' high-density property makes intersections highly coupled with each other. All these characteristics make it crucial to consider future uncertainties. Hence, rolling optimization technologies got noticed by traffic researchers and have been developed for multiple decades.

Based on how the problem is decomposed, rolling optimization can be done with either centralized, decentralized, or hierarchical paradigms. Rolling optimization ATSCs also adopt various mathematical tools such as dynamic programming (DP) (Chen and Sun 2016; Yao et al. 2019; Chiou 2019), approximate DP (Cai, Wong, and Heydecker 2009), mixed integer linear programming (MILP) (Guilliard et al. 2016), and so on. Limited by the space, we only discuss some representative approaches below.

1. Third generation (3G) ATSCs developed before the 21st century broadly use the DP algorithm to solve a decentralized rolling optimization problem. Different from 2G ATSCs, this generation's approaches adjust signal plans continuously, rather than based on the time of the day. Some of them even explore the possibility of using the VPS actions space for higher flexibility. In-field studies evaluate their real-world applications and report the performance improvement in criteria like traffic delay compared to actuated controllers. 3G ATSCs push the field forward with their methodology innovation but still suffer from some limitations. Restricted by the capabilities of sensing, communicating, and computing at that time, 3G methods rely heavily on loop detectors (expensive and inefficient), compromise relatively simple traffic dynamic models, and may become intractable in network-wide applications. Some examples from this generation are OPAC (Gartner 1983), PRODYN (Henry, Farges, and Tuffal 1984; Farges, Khoudour, and Lesort 1994), UTOPIA (Mauro and Di Taranto 1990), and RHODES (Head, Mirchandani, and Sheppard 1992; Mirchandani and Head 2001).

2. Successors, also denoted as the fourth-generation ATSC in (Gartner, Stamatiadis, and Tarnoff 1995), dig deeper into both the aforementioned shortcomings and optimization algorithms. Many articles assume similar settings to the 3G ATSCs and explore the possibility of using other dynamic models and optimizers. SURTRAC (Xie et al. 2012; Flow, n.d.), a decentralized method with neighboring communication, describes the problem as a scheduling system assuming platoons are indivisible for simplifying the model. The model leads to a MILP problem and reduces the computation complexity. It is deployed and evaluated in Pittsburgh, PA, USA. A study report indicates its effectiveness (Smith et al. 2013). Similarly, (Guilliard et al. 2016) builds a queue-transmission model endorsing heterogeneous prediction time-scope for computation simplicity. Model predictive control (MPC) methods in the control field, as a special case of rolling optimization methods, have been introduced to the ATSC problem for more than two decades (Ye et al. 2019). Ideally, these control theory-based methods bring more reliability since most of them come up with the proof of stability, but none of them has been tested in the field due to the similar issue 3G ATSCs encountered.

**Model-free reinforcement learning**

Considering the limitations rolling optimization approaches have, especially the accessibility and reliability of dynamical models, researchers in the early 2000 turned to seek model-free and low online computational complexity methodologies. The rapid developing model-free reinforcement learning (MF-RL) technique met both requirements and became a popular topic in the ATSC field till today. It works under a similar objective to the rolling optimization approaches — optimizing the expected future reward. But they work in different ways. From the decision-making side, MF-RL turns the online signal plan optimization problem into an offline policy optimization problem by leveraging its self-learning property. The learned policy significantly reduces online computational complexity. From the optimization side, MF-RL estimates the expected future return directly from experiences collected through interaction with the real system or simulators without modeling the system dynamics. These features help MF-RL ATSCs overcome the limitations RHO-based ATSCs facing.

(Abdulhai et al 2003) and (Abdulhai and Kattan 2003) first introduced the potential of adopting RL in ATSC problems. Multiagent RL for integrated network (MARLIN) (El-Tantawy, Abdulhai, and Abdelgawad 2013) then extended the discussion to a multi-agent coordination setting. MARLIN applies the decentralized paradigm, designs neighboring communication schemes, and exploits intersection-level coordination by leveraging game theory methodologies. It is one of the few RL-based ATSC methods ever tested in the field.

The classic Q-learning that the MARLIN system uses is proved to converge (Melo 2001) under the assumption of fully Markov and stationary (Baird 1995). However, its tabular RL form restricts the representative capability of learned policies. Too coarse state and action space definition may adversely affect accuracy; while too fine-grained space definition increases the look-up table exponentially and makes it intractable. This dilemma was alleviated by the function approximation and deep learning technologies. Combining them together, deep reinforcement learning (DRL) conquers many classic planning and control tasks and even receives public attention (Silver et al. 2016; Vinyals et al. 2019; Mnih et al. 2015). By introducing DRLs into the ATSC field, many promising solutions are proposed (Wei, Zheng, et al. 2019; Wei, Chen, et al. 2019) and waiting to be examined by in-field tests.

Deep neural networks (DNNs) as the function approximators grant DRL the capability of handling high-dimension observation input, which takes advantage of merging sensing techniques and contains richer information. (Genders and Razavi 2016; Wei et al. 2018; Alizadeh and Abdulhai 2018; Tan et al. 2019) adopts the deep Q-learning (DQN) with convolution neural network (CNN) approximators. They work under the assumption that traffic surveillance cameras, radar, LiDar or connected vehicles have been promoted and detailed vehicle states can be accessed in real-time. Many recent generic DRL research explores the direction of centralized training decentralized executing (CTDE) framework enabled coordinated multi-agent control. CTDE is a multi-agent RL regime that grants agents access to global information in the learning stage (interacting with the simulator). The observation space each agent making decisions from is the same as in the deployment stage. But the policy performance and the agent learning can be evaluated and guided by the global knowledge, which encourages policies contributing to the global optimum better. Once the policies are fixed and deployed in-field, agents only follow their local policies and do not require the global information any longer. Such a structure fully utilizes the plentiful resource in the virtual environment and respects the real-world decentralized deployment's requirements in the meantime. Pioneer CTDE research is value decomposition network (VDN) (Sunehag et al. 2018), counterfactual multi-agent policy gradients (COMA) (Foerster et al. 2018), Q-Mix (Rashid et al. 2018), Q-TRAN (Son et al. 2019), and ROMA (Wang et al. 2020). When introducing the CTDE structure into the ATSC field, it is combined well with graph-based structural learning methods because of the nature of traffic network topology. For example, graph attention networks (GATs) are used to capture spatial relationships between neighboring intersections (Nishi et al. 2018; Wei, Xu, et al. 2019; Devailly, Larocque, and Charlin 2021).

The MF-RL approach is promising, but cannot be perfect. First of all, Markov and stationarity assumptions that the classic Q-learning requires cannot be always met. The function approximation will further introduce the estimation error (Baird 1995). As a consequence, DRLs are not guaranteed to find the optimum. Secondly, the feature that MF-RLs do not attempt to model the system dynamic poses a great burden on the agent-environment interaction data, to evaluate and improve the value function or policy properly. More importantly, the learned policy optimized on the interaction samples may not generalize over vastly different scenarios since the sample-implied system transition distribution may differ from the real one. In the case that the distribution shifting occurs — traffic demand and neighboring intersections behave differently from the training scenarios — the stability of MF-RL-based ATSCs cannot be ensured. These common issues in the generic MF-RL field have not attracted traffic researchers' attention till very recently. (Jaggi et al 2021) has shown that MF-RL may not generalize well if demand patterns shift considerably from the training demand. They introduced Monte Carlo Lights: a first ATSC system that explores model-based RL and Monte Carlo Tree Search in the field of traffic. Most of the recent RL ATSC literature, however, does not focus on generalization, robustness, and stabilizing capabilities. (Devailly, Larocque, and Charlin 2021) is one of the very recent studies that considers the generalization issue and claims that graph convolution networks (GCNs) can generalize well with enough interaction samples from more training scenarios. But these issues remain as emerging research topics.

### III. A Unified View for the ATSC Problem: Problem Modeling

As discussed in Section II.B, the TSC problem can be handled in different ways, including but not limited to control system, queuing system, scheduling problem, etc.

Over the past few decades, particularly in the last few years, contributions to the ATSC field have been abundant and rapidly expanding. Most recently, and most relevant to this chapter, there has been an explosion of publications on multi-agent deep RL approaches to ATSC. Common to all are: (1) a choice of RL method; value-based or policy-based, (2) deep neural network architecture such as variants of CNNs and GNNs, and (3) a choice of state and reward definition. It is challenging to benchmark the variety of such methods relative to each other, let alone advance them or compare them to classical methods such as model predictive control methods.

Therefore, we believe there is a significant technical gap in the field, which leads to the major task of this article: to develop a novel unified view of modern ATSCs in order to identify common grounds as well as differences and shortcomings of existing methodologies with the ultimate goal to facilitate cross-fertilization and advance the state-of-the-art in ATSC. We argue that, although the existing approaches have major and subtle differences, they can be cast as the same mathematical problem and be viewed as solutions with different mathematical methodologies. In this section, we introduce our proposed unified view and the necessary preliminaries, pose the TSC problem as the widely adopted Markov Decision Process (MDP), and provide a unified view upon the different fields of methodologies to better understand their differences and relative contributions.

After establishing the unified view, which is our first objective, we then revisit the existing ATSC approaches with the established unified view to specify our full set of ATSC research objectives in Section IV and present a methodological path forward in Section V that will hopefully contribute to guiding future developments in this domain.

*A. Preliminaries*

*1) Stochastic Processes and Dynamical Systems*

**Stochastic processes**

A stochastic process $\{X_t\}, t \in T$ describes a family of random variables indexed by $t$. The index $t$ can be either continuous (e.g. continuous-time Wiener process, $t \geq 0 \in \mathbb{R}^+$) or discrete (e.g. toss coins multiple times, $t = 0,1,...,n,... \in \mathbb{N}$). In this article, we mainly focus on the discrete case, in which we discretize the time into time steps: $\{X_t\}, t \in \mathbb{N}$.

Stochastic processes as a mathematical model are widely adopted across various fields with the capability of modeling generic correlations among ordered events (the order indexed by t): the random variable distribution can either be history-dependent or independent. A simple ball drawing example shows the difference. Assume there are 2 red balls and 1 yellow ball in a bag. We draw balls with replacement from the bag for three times. The distributions of getting colors of balls every time are then independent and identical since we always have the same candidate balls. This gives us a process in which $P(X_t), t = 1,2,3$ are independent and identical distributed (i.i.d.).

In the same experiment, changing the setting from "with replacement" to "without replacement" gives us a history-dependent stochastic process since in time step 2 and 3, the candidate ball sets are determined by the happened event in previous time step $t = 1$, etc. Then we get conditioned distributions $P(X_1), P(X_2|X_1), P(X_3|X_1, X_2)$. Clearly, the history-independent stochastic process can be treated as an extreme case of the second history-dependent stochastic process. We denote the latter as a generic representation of stochastic processes.

**Dynamical systems**

Dynamical systems are a methodology describing systems' time-evolution by capturing their "key states". A system's state is commonly a vector of real numbers that lie in the state space $\mathbf{x}_t \in \mathcal{X} \subset \mathbb{R}^n$. Dynamical systems can be defined in either continuous or discrete time. Correspondingly, ordinary differential equations (ODEs) and difference equations as transfer functions describe the evolution of the system state and imply time dependency. For example, the simplest discrete-time linear system defines an affine mapping and implies the one-step history dependency:

$$\mathbf{x}_{t+1} = A \cdot \mathbf{x}_t + \mathbf{b}, \tag{1}$$

where matrix A and vector b are constant-coefficient matrices manifesting the system behavior.

More generally, the transition of a dynamic system can be history-dependent and stochastic as well. And this is where dynamical systems meet stochastic processes: Let's consider the state vector $\mathbf{x}_t$ as a sample of a stochastic event in the sample space, $X_t = \mathbf{x}_t$, and extend the affine mapping in (1) to a generic form, $f: \mathcal{X} \mapsto \mathcal{X}$, where the mapping is no longer deterministic and is represented by conditional probabilities $P(X_{t+1}|X_t)$. This generalization yields the fact that dynamical systems can be represented by stochastic processes.

Formally defining stochastic processes and dynamical systems requires many more mathematical tools. And mathematicians paid their efforts in bridging the two concepts theoretically. In order to keep this article concise, this section only describes their relationship intuitively. For those who are interested in the detail, we refer readers to article (Collet 2008).

**A first glance at the control — the structure of control systems**

Control theory is a well-known research discipline using dynamical systems. And many traditional methods developed from there had been adopted to the traffic field (Papageorgiou et al. 1991; Diakaki, Papageorgiou, and Aboudolas 2002; Kotsialos et al. 2002), not limited to TSCs. Hence, it is critical to have a general understanding of this branch in order to build our unified view.

Control theory builds a methodology for designing controllers manipulating systems to desired states through governing the system input. The controller design targets not only optimizing designated criteria like steady-state error, which is similar to an optimization problem, but also on ensuring a certain degree of system stability. The stability may refer to different definitions in various scenarios, such as the bounded-input bounded-output stability and Lyapunov stability, which is beyond the scope of this article. While we are more interested in how they form and analyze the system. Hence, we briefly define basic concepts in a control system, together with their corresponding terminologies in operations research (OR) (optional, in parentheses) which will be dominantly used later:

- Controlled system (the state space): the subject to be controlled, e.g., a plant, process, traffic intersection.
- Controller (agent): the executor generating control (action) following a human-designed control law.
- Control law (policy): the mapping from any prerequisites to the control (action).

- System input: the input signal driving the dynamical system evolution. It is not purely the control signal, but a combined signal after applying to the original input — the desired/reference value, noise, and the outer-system influences (see Section III.A.4).

The "prerequisites" of the control law is not limited to the system state (system outputs sometimes can be considered as a part of the system state, for example, the room temperature within an air conditioning system) and can be various: the original system inputs (the target value), or even just the time of the day. The type of prerequisites and dependencies identifies the three structures of control systems: open-loop control, closed-loop (feedback) control, and feedforward control. In the air conditioning system example:

A control law only conditioned on the time of the day, i.e., higher cooling level around noon, lower after sunset, and neglecting the current indoor temperature forms an open-loop control system.

A control law conditioned on the current indoor temperature forms a feedback control system.

A control law considering the current outdoor temperature, which would influence the indoor temperature, and adjusting the cooling power according to it forms a feedforward control system.

Fig 1.4 illustrates three structures' workflow with block diagrams. One may argue that, ideally, the "feedforward" branch is unnecessary and can be annihilated by including the outdoor temperature into the system state, which forms a state-feedback structure. However, in most engineering applications, this is intractable. Section III.A.4 provides the counterarguments and detailed discussion.

The open-loop control law is commonly derived from historical observations, e.g., the historical temperature trajectory, and does not react to any real-time system variations. It is the naivest approach and lacks guarantees of either stability or optimality. The feedback structure feeds the controllers with real-time system information. The key objective of control theory — the stability — relies on this closed-loop structure which grants controllers the ability to react to and adjust the system in real time. A feedback control system can be proved to have stability through careful control law design, and hence became the most broadly used structure. A feedback controller ensuring the system stability may even not require the knowledge of the system dynamics: the proportional-integral-differential (PID) method, which is adopted in the ALINEA (Papageorgiou et al. 1991) ramp metering control algorithm.

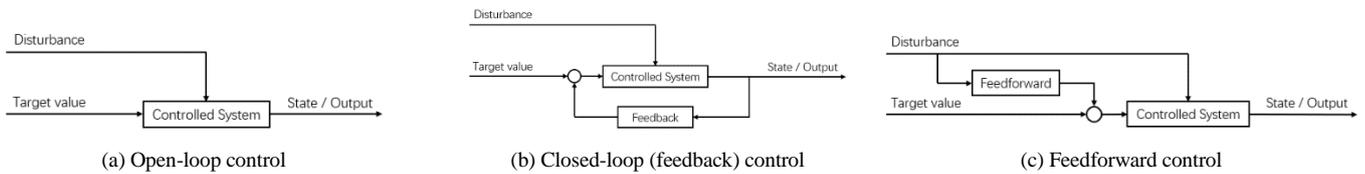

(a) Open-loop control     (b) Closed-loop (feedback) control     (c) Feedforward control

Figure 1.4. Three basic control structures with block diagram representation, where the circle represents the combination of signals. The disturbance refers to the outer-system influence. The target value refers to the set point in process control and motion control, which can sometimes be ignored in optimal control scenarios.

The feedback structure performs well in traditional scenarios like process control and motion control, where the aim is to drive the system to the reference value. While, in more complex scenarios, systems may be expected to follow a trajectory (e.g., ballistic control), or will be impacted by outside disturbances significantly (e.g., some optimal control cases), in which case only the feedback signal is not sufficient for reaching the objectives. Either the target trajectory or the disturbance is about the future. Hence, a control law conditioned on and activating according to these prerequisites is considered as "feedforward". A pure feedforward control can be treated as an extreme case of open-loop control since it is not conditioned on the system state. But in most cases, feedforward control is used in combination with feedback, for example, the motion control (Meckl and Kinceler 1994) and the model predictive control (MPC), see Section IV.A.3.

**Linking control systems to stochastic processes**

At last, in this part, we show how to convert a special and simple linear control system into a stochastic process with an example. Notably, we cannot always do this for all control systems considering the variety of control laws. Specifically, the optimal control problem this chapter facing (TSCs) meets an extension of stochastic processes, i.e., the Markov decision processes (MDPs), which is covered in Section III.A.3. This part only illustrates the connection from an intuitive perspective.

We consider a pure feedback, discrete-time controller with a linear system in the state-space representation:

$$\begin{aligned} \mathbf{x}_{t+1} &= A \cdot \mathbf{x_t} + B \cdot \mathbf{u_t}, \\ \mathbf{u_t} &= -K \cdot \mathbf{x}_t, \end{aligned} \qquad (2)$$

where $\mathbf{u_t}$ is the state-feedback control signal, and matrix $K$ identifies the control law (sometimes called the gain). Combine the above two equations, we have the dynamical model of the whole control system:

$$\mathbf{x}_{t+1} = (A - BK) \cdot \mathbf{x}_t. \qquad (3)$$

It has the same form as (1). And it hence can be treated as a discrete-time stochastic process. Despite (2) shows a very simple system, it can even express a significant optimal control method — the linear quadratic regulator (LQR), when the problem doesn't have constraints.

*2) Stochastic Processes and Dynamical Systems: Properties*

For unifying the language, from now on, we use $s_t$ to represent an event in a stochastic process and a state in a dynamical system, instead of $x_t$. Also, we restrict most discussions to discrete cases for simplicity.

Using probabilistic graphical models (PGMs), a generic stochastic process can be expressed as in Fig 1.5(a), where arrows indicate the dependency between two events. As we stated before, such a representation can express the generic form of dependency on the full history and is capable of modeling any processes. However, in real-world applications, the full dependency makes both the process identification and inference sophisticated. We need a simplification.

If a system's state definition provides *enough information* for predicting the next state based on only the present state, we can neglect most of the arrows in Fig 1.5(a) and keep the one-step dependencies only, as shown in Fig 1.5(b). Such a process meets the Markov property and is named a discrete-time Markov process (a finite-discrete-state space defines the famous Markov chain).

**Definition 1 (Discrete Markov Process)** *A discrete process $\{s_t\}$ is a Markov process if its state transition distribution meets the Markov property, that is, the distribution of a new state is solely determined by the present state:*

$$\Pr(s_{t+1}|s_0, s_1, \ldots, s_t) = \Pr(s_{t+1}|s_t), \forall t \in \mathbb{N}. \tag{4}$$

It is worth to note that we can convert a generic discrete-time stochastic process into a Markov process by defining the state space carefully. For a non-Markov process whose transition is described by $\Pr(s_t|s_0, s_1, \ldots, s_{t-1})$, defining $H_t = (s_0, s_1, \ldots, s_t)$ gives us a new Markov process $\{H_t\}, t \in \mathbb{N}$.

Even if a dynamical system is fully Markov and discrete-in-time, it is still too complicated to model state transitions for each time step. Hence, we hold another assumption to further simplify the model. Homogeneity assumes the transition distribution keeping invariant across the time horizon. It grants us the opportunity to model the system with a single conditional distribution. The PGM representation then is simplified to Fig 1.5(c), where s and s' represent the present and next events respectively. Formally:

**Definition 2 (Time-Homogeneous Markov Process)** *A discrete Markov process $\{s_t\}$ is homogeneous if its state transition distribution meets the homogeneity, that is, the conditional distribution describing the system transition does not condition on the index t:*

$$\Pr(s_{t+1}|s_t) = \Pr(s'|s), \forall t \in \mathbb{N}. \tag{5}$$

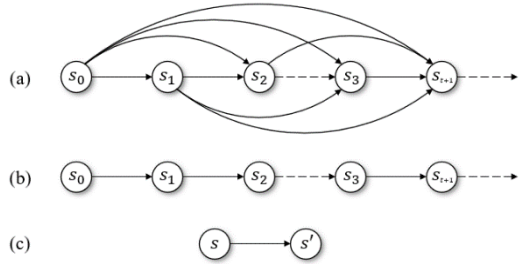

Figure 1.5. An illustration of discrete stochastic processes: (a) a general stochastic process, (b) a Markov process, (c) a homogeneous Markov process

Different from the Markov property, it is only possible but not guaranteed to convert a non-homogeneous process to a homogeneous one by re-defining the state representation. For example, performing differencing on consecutive states may stabilize some non-stationary processes.

This time homogeneity has different names in various research fields. It is called the "time-invariant" property in the control field and the "stationarity" in MDPs. While, in the context of the stochastic process, the stationarity refers to other definitions, that readers shouldn't confuse.

**Definition 3 (Strict-Sense Stationary Process)** *Let $F_S(s_{t_1}, \ldots, s_{t_n})$ represent the cumulative distribution function of the joint distribution of events $s_{t_1}, \ldots, s_{t_n}$. In a generic continuous-time setting, a process $\{s_t\}$ is strictly stationary if $F_S(\cdot)$ is not conditioned on the index t. Formally:*

$$F_S(s_{t_1+\tau}, \ldots, s_{t_n+\tau}) = F_S(s_{t_1}, \ldots, s_{t_n}), \forall \tau \in \mathbb{R}, \forall t_i \in \mathbb{R}^+, and\ \forall n \in \mathbb{N}. \tag{6}$$

Many processes like the white noise and the with-replacement ball drawing game mentioned earlier have strict-sense stationarity. Obviously, it is stricter than the time-homogeneity a Markov process may have. Another definition, the wide-sense stationarity, requires a process having a time-invariant 1-st moment (mean) and autocovariance, and finite 2-nd moment (variance). It imposes weaker constraints but is still stronger than time-homogeneity. In order to eliminate ambiguity, in the following MDP context, stationarity inherits *the weakest definition*, the time-homogeneity.

*3) Markov Decision Processes (MDPs)*

MDPs extend discrete Markov processes considering the impact of the decision-maker and is a mathematical framework for modeling sequential decision-making problems broadly used in many disciplines. In every time step of an MDP, the decision maker selects an action, moves the system state to another according to its dynamic, and receives an immediate reward. From the perspective of a control system, it forms a feedback loop. The goal of solving an MDP problem is to maximize the expected total reward over the future at each decision point.

**Stationary infinite-horizon MDPs**

For simplicity, we first consider a stationary process. An infinite-horizon MDP is a tuple $\langle \mathcal{S}, \mathcal{A}, P, R, \gamma \rangle$, where $\mathcal{S}$ is the state space; $\mathcal{A}$ is the action space; $P(s'|s,a)$ is the distribution of the system transition; $R(s,a)$ is the immediate reward function; $\gamma \in [0,1]$ is the discount factor. A stationary, infinite-horizon MDP can be illustrated with a graphical model as shown in Fig 1.6, where circles represent states; squares refer to actions; diamonds refer to utilities (rewards, values, etc.). Under this setting, we refine the definition of policy given in Section II.B.2. A policy $\pi(a|s) \in \mathcal{P}(\mathcal{A})$ is a mapping from state space to action space. The mapping can be either probabilistic or deterministic. All distributions $P, R, \pi$ are assumed to be stationary, which means they do not vary through time. The expected discounted total reward under a policy $\pi$ of each state is defined as the value function $V^\pi(s)$:

$$V^\pi(s) = \mathbb{E}\left[\sum_{t=0}^{\infty} \gamma^t R(s_t, a_t)\right]. \tag{7}$$

By splitting the immediate return term out, we get its recursion form, also known as the Bellman equation:

$$V^\pi(s) = \mathbb{E}_{a \sim \pi}[R(s,a)] + \gamma \sum_{s' \in \mathcal{S}} P^\pi(s'|s) V^\pi(s'). \tag{8}$$

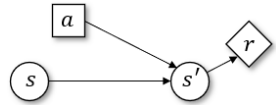

Figure 1.6. A stationary, infinite-horizon Markov decision process

To solve this MDP, we find an optimal policy $\pi^*$ that maximizes the value function over the state space, which gives us the optimality equation:

$$V^*(s) = \max_\pi V^\pi(s) = \max_{a \in \mathcal{A}} \left\{ R(s,a) + \gamma \sum_{s' \in \mathcal{S}} P(s'|s,a) V^{\pi^*}(s') \right\}, \forall s \in \mathcal{S}. \tag{9}$$

**Example 1** *Consider an isolated intersection TSC problem while taking traffic surveillance cameras as traffic sensors (see Section II.A.1, state perception part). As shown in Fig 1.7, the red polygon identifies the controlled intersection's detection area surveillance cameras could cover. Make the following assumptions: 1. Assume that we can precisely measure necessary system states from surveillance cameras' output, which guarantees the completeness of the modeled system. 2. Assume that the intersection's traffic dynamic under each action is stationary. 3. Assume that the traffic demand on all incoming arms is time-invariant, that is, the influence upon the controlled intersection's dynamic from upstream intersections and peripheral traffic network is time-invariant and guarantees the system's stationarity. Let s, a, r represent state, action, and immediate reward of the intersection. Then we can formulate this problem as a classical stationary MDP as shown in Fig 1.6.*

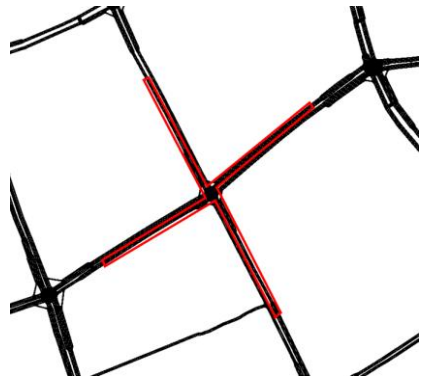

Figure 1.7. Control an isolated intersection within a traffic network

We can find a policy that selects the best action according to the current state directly by solving such an MDP. That is, the optimal policy class of MDPs is the *state-dependent deterministic policy*.

Non-stationary infinite horizon MDPs are extremely complex since all components are time-variant which leads the Bellman backup in (8) commonly intractable.

**Factored Markov Decision Processes (FMDPs)**

The optimization problem given by (9) can be solved as a whole while the scale of the system is acceptable (state and action spaces). But it becomes intractable while modeling larger-scale systems, such as multi-robot or multi-intersection TSC problems. Factored MDPs (Guestrin, Koller, and Parr 2001; Guestrin et al. 2003) provides a compact and structured representation to the system transition by using dynamic Bayesian networks (DBNs), which is a typical probabilistic graphical model (PGM) in Bayesian statistics. FMDPs decompose the full system state into components, decouple the relationships among state components, and abandon unnecessary links which will all be included by classical MDPs. Hence, the DBN simplifies and clarifies the system representation.

In a FMDP, the full system state is described by a vector of states $\mathbf{s} = [s_1, \ldots, s_n]^\top$, $\mathbf{s} \in \mathcal{S} = \mathcal{S}_1 \times \mathcal{S}_2 \times \ldots \times \mathcal{S}_n$. Same for the full action $\mathbf{a} = [a_1, \ldots, a_m]^\top$. The single transition function $P$ is replaced by factors $P_i(s'_i | \mathbf{s}_j, \mathbf{a}_k), \mathbf{s}_j \in \mathcal{S}_j \subseteq \mathcal{S}, \mathbf{a}_k \in \mathcal{A}_k \subseteq \mathcal{A}, \forall i$ denoting partial system transitions for state $s_i$, where $\mathbf{s}_j$ and $\mathbf{a}_k$ are the sets of variables influencing $s_i$ directly. Similarly, the reward function $R$ can be decomposed as well $R = f(R_1(s_1), \ldots, R_n(s_n))$, where $f(\cdot)$ is the aggregation function providing the global reward, for example, the summation function. We then show an example of a FMDP with a simple traffic network.

**Example 2** *Consider controlling a corridor consists of three intersections as shown in Fig 1.8(a). We still use surveillance cameras as sensors. And each intersection has a detection area. Let $s_1, s_2, s_3$ and $a_1, a_2, a_3$ represent local state and action spaces of intersection 1,2,3 respectively. In each time step, we count the queue length $Q_i$ of each intersection $i$ as their immediate reward $R_i(s_i) = Q_i$. Inherit assumptions 1-3 from Example 1 and further assume that each intersection's action only affects the immediate next state of the corresponding intersection and its state influences neighbors' future. We sum local rewards together as the global reward $R(s) = \sum_i R_i$. Then we can formulate this problem as an FMDP as shown in Fig 1.8(b). Notice that how the graphical representation displays the relationships among state and action components, and how is the value function factorized into value components $h_i$. For example, the factor denoting the transition of $s_1$ is $P_1(s_1' | s_1, s_2, a_1)$.*

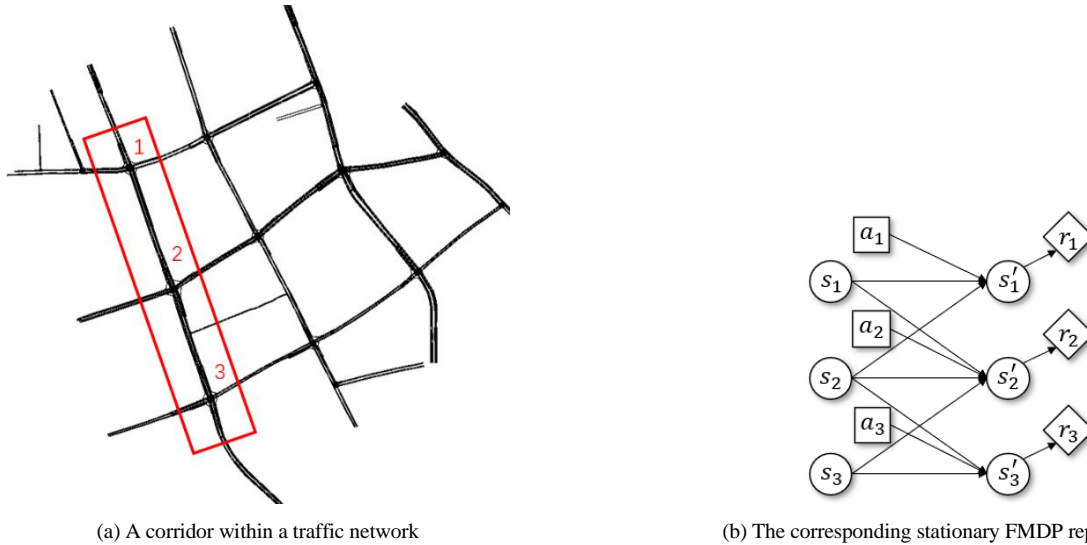

(a) A corridor within a traffic network      (b) The corresponding stationary FMDP representation

Figure 1.8. Formulate a simplified three-intersection corridor ATSC problem with a stationary FMDP

Of course, we can form such a corridor-control problem as a standard MDP by joining state spaces, action spaces, and rewards, and present it by Fig 1.6. However, by doing this, we are linking all state and action components together. Taking intersection 1 as an example again, in an MDP representation, its local state transition is dependent on global joint state-action, $P(s_1' | s_1, s_2, s_3, a_1, a_2, a_3)$. This is an example of the curse of dimensionality that the generic MDP suffers from as the size of the problem grows exponentially with the number of intersections. In contrast, FMDP alleviates such a curse of dimensionality. Its factor $P_1$ in Example 2 largely simplifies the model.

It is worth noting that, for an FMDP, we still aim to find a joint policy for solving the whole problem rather than independent policies. The correlation between local policies here is given by the aggregation of value components $h_i$. Hence, the optimal policy class of FMDPs is still the *state-dependent deterministic joint policy*.

**Partially Observable Markov Decision Processes (POMDPs)**

In the above MDPs, we potentially hold another assumption that all system states can be explicitly observed. Those MDPs are also called fully observable MDPs. But this cannot be always met in real-world problems. True states can be polluted by noise and measurement error. Some of them even cannot be measured caused by the limited sensing capability or the nature of states: loop detectors, being point detectors, restrict the observability of the system state (e.g. full queue) and the observability of the upcoming demand, which can only be possibly predicted rather than measured. Hence, in most of the cases, (9) cannot be optimized directly.

POMDPs (Kaelbling, Littman, and Cassandra 1998; Roy, Gordon, and Thrun 2005) are introduced to describe such a partial observability. A POMDP extends the MDP tuple with two extra elements $\langle S, \mathcal{A}, P, R, \Omega, O, \gamma \rangle$: the observation space $\Omega$, and the conditional observation function $O(o|s)$. Fig 1.9 shows a stationary POMDP, where dashed-circles represent partially observable states; and arrows directing from dashed-circles to solid-circles are the conditional observation functions. We extend the traffic Example 1 by relaxing some assumptions to illustrate a POMDP.

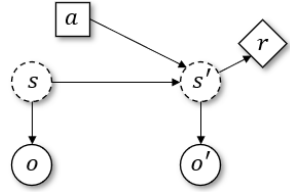

Figure 1.9. A stationary POMDP

**Example 3** *Consider the same isolated intersection TSC problem. Inherit assumptions 2 and 3 from Example 1 but exclude the first one. Assume the only sensor in the field is the loop detector, which gives us partial information about the system state. Let $s$, $o$, $a$, $r$ represent true state, loop detector readings, action, and immediate reward respectively. Then we can formulate this problem as a POMDP as shown in Fig 1.9. Notice that in such a process, the transition between consecutive observations is not directly linked. Hence, we cannot assume Markov and stationary properties on $P(o'|o)$ as we did on the state anymore.*

We can't optimize a present-observation-dependent policy on such a process. Solving a POMDP requires full knowledge of the observation-action history since the evolving of partial observation is neither Markov nor stationary. But, considering the complexity, it is also not practical to optimize a full-history-dependent policy. We commonly estimate a belief state $b \in \mathcal{B}$, as a probability distribution over the true state space $S$ and cast the POMDP into a fully observable belief MDP problem for a simpler solution. A POMDP policy $\pi: \mathcal{B} \mapsto \mathcal{A}$ then maps the belief, rather than the state, to the action. Correspondingly, let the reward given a belief as an expectation over the state and define the value function and Bellman equation:

$$V^\pi(b) = \mathbb{E}\left[\sum_{t=0}^{\infty} \gamma^t R(b_t, a_t)\right] = \mathbb{E}\left[\sum_{t=0}^{\infty} \gamma^t \left[\sum_{s \in S} b(s) R(s, a_t)\right]\right], \tag{10}$$

$$V^*(b) = \max_\pi V^\pi(b) = \max_{a \in \mathcal{A}} \left\{ R(b, a) + \gamma \sum_{o' \in O} P(o'|b, a) V^{\pi^*}(b') \right\}, \forall b \in \mathcal{B}, \tag{11}$$

where $b'$ is the updated belief updating recursively given previous belief $b$, action $a$, and new observation $o'$. The derived belief MDP share the same optimal policy class as MDPs: *belief-dependent deterministic policy*. While an origin POMDP has an *observation-history-dependent deterministic optimal policy*.

**Mixed Observability Markov Decision Processes (MOMDPs)**

When a system contains both fully and partially observable information, it is too complicated to solve a pure POMDP since maintaining the belief over full states is expensive. By decomposing the system state with FMDPs' idea, we can get an MDP with different observable state components, which is called mixed observability MDPs. It can be seen as a factored POMDP where each state component $s_i \in S_i$ could have its own observation function $O_i(o_i|s_i)$. Different observability gives various observation functions from null observable $O_i(\cdot): S_i \mapsto \emptyset$ to full observable $O_i(\cdot): S_i \mapsto S_i, O_i(s_i) = s_i$.

Fig 1.10 represents a simple MOMDP with two state components $S = \mathcal{X} \times \mathcal{Y}$, where $x$ is fully observable; $y$ is partially observable. The observation contains two parts as well: $\mathbf{o} = (o_x, o_y) = (x, O_y(y))$, where the $x$ state component is fully observable and its corresponding observation function is an identity mapping $O_x(x) = x$. Notice that rounded rectangles in Fig 1.10 represent the collections of state (observation) components rather than the "plate notation" representing repeating variables used in Bayesian inference (Buntine 1994, 174–77). And this state collection notation keeps the same in the following context.

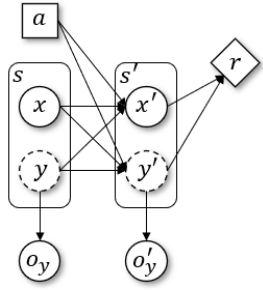

Figure 1.10. A stationary MOMDP

MOMDPs introduce a compact representation and simplify the solution (Araya-López et al. 2010; Ong et al. 2010). We only keep updating the belief estimation of the non-fully observable part of the system $\mathbf{b} = (b_x, b_y) = (x, b_y)$, rather than the full system. A MOMDP as a factorized extension to POMDPs holds the same optimal policy class, that is the *observation-history-dependent deterministic policy*. But MOMDPs' optimal policies are easier to either search or execute (need lesser effort on online belief updating).

*4) System Boundary and Exogenous Events*

The above introduced stationary MDPs are powerful and efficient to model "closed systems", whose dynamics will not be affected by disturbances from the outside. Classical applications potentially hold such an assumption. Taking maze-searching as an example, we typically do not consider the case that stochastic events block some routes. Even in the scenarios that we have to take these events into account, MDPs can easily handle the modeling problem by extending the state space to including those events (Boutilier, Dean, and Hanks 1999).

From the modeling perspective, we name the behavior of determining state spaces as system boundary delimiting, that separates *inner system* and *outer system*. The impact from outside to the modeled system is called *exogenous events* (Boutilier, Dean, and Hanks 1999). How to delimit the boundary for MDPs modeling solutions becomes a major task for ATSC designers. It is commonly not realistic to fully model a problem, for two reasons: 1. From a systematic viewpoint, external disturbance is widespread. As in traffic, an urban traffic network is ubiquitously connected to freeways, suburbs, and other modes of transportation. And its dynamic is affected by non-traffic events such as the weather, which can be observed but is hard to be modeled. 2. Modeling a large system in an MDP solution is expensive and uneconomical. As in TSC, a vehicle driving from Mississauga to Toronto may finally affect the traffic state in downtown Toronto. But modeling such a vehicle when it is still in Mississauga is unreasonable.

Precisely, exogenous events stochastically affect the system and move the system state to another. It should not be confused with the agent's action which causes state transition as well since the exogenous event is beyond the control of the decision-maker. For example, we consider a humanoid robot as a system and its posture as the full state. When a research staff pushes the robot, its state transits while there is no action has been taken.

Exogenous events correspond to the evolution of the natural process beyond the inner system. In ATSC solutions, when we are modeling a sub-network, the peripheral network's dynamic is omitted but still evolving. Then the traffic demand coming from the peripheral network is the exogenous event to the sub-network. On the other hand, in a decentralized TSC system, each agent is assigned to model an intersection or a sub-region. Actions from one agent's neighbors affect this agent's state but are not captured by its model, hence they are exogenous events as well. We discuss how to model the TSC's world with MDPs and how to consider the exogenous events as a part of an MDP next and leave the solution discussion to Section IV.A and Section IV.B.

*B. ATSCs: A MOMDP Representation*

In the simplest case, we divide the system state into two components by the selected system boundary, and model TSCs as MOMDPs. $\mathbf{s} = [s_c, s_o]^\top \in \mathcal{S}$, where $s_c$ represents the controlled inner system; $s_o$ is the observed-only outer system state. For simplicity, we presume that $s_c$ can be explicitly measured and estimated. The peripheral network $s_o$ mainly affects the inner system in the way of traffic demand. It is natural and feasible to measure and predict the traffic demand as an observation to the $s_o$ on the delimited system boundary. Letting $d$ represents the demand on the boundary, $o = (o_c, o_o) = (s_c, O_o(s_o)) = (s_c, d)$. In this case, $d$ is the exogenous event influencing the inner system directly. An example shown below explains it in detail.

**Example 4** *Consider controlling the same isolated intersection introduced in Example 1. Let the red polygon in Fig 1.7 identify the detection area also denotes the system boundary. Inherit assumptions 1 and 2 from Example 1, exclude the third one, and make extra assumptions: 1. Assume that the outer system evolves stationary and only affects the inner system by posing traffic demand. 2. Assume that loop detectors are deployed at the end of the detection area for estimating the incoming traffic demand. Let $s_c$, $s_o$, $d$ represent the intersection state, peripheral traffic network state, and the traffic demand flowing to the controlled intersection respectively. Then, the MOMDP can be formed as in Fig 1.11. It is different from the generic MOMDP shown in Fig 1.10, since that the action and rewards only relate to the controlled part, namely, the inner system. This structural difference gives us an opportunity to find an approximated optimal solution to the inner system: optimizing the whole world is not reachable.*

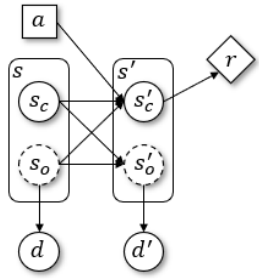

Figure 1.11. Stationary MOMDP representation of an isolated ATSC problem considering demand as exogenous events

Example 4 delimits the system boundary fixed to the controlled area. In corresponding solutions, the non-fully-modeled exogenous events affect the controlled area directly in time. As we discussed earlier in this section, by extending the system boundary, MDPs can do better. Despite we will not model the whole outer system, expanding boundary helps to alleviate the impact the new outer system poses on the controlled area. Back to the Mississauga vehicle example, to control a group of intersections in downtown Toronto, it is necessary to perceive the traffic statistics around the controlled area (e.g., few-hop neighboring intersections), but there is no need to observe a vehicle that is leaving Mississauga. That vehicle can be treated as white noise to the inner system and be ignored — the geographical distance defers and alleviates its impact. Hence, we further extend the MOMDP model to a three-stage factored MOMDP model that covers generic TSC scenarios.

**Example 5** *We still control the same isolated intersection introduced in Example 1. In this case, besides the local traffic state, presume that we can acquire some neighboring intersections' information as well. As shown in Fig 1.12(a), the red rectangle represents the controlled intersection; the blue polygon indicates the observed and modeled neighboring area, and the **new system boundary**; the green dashed-rectangle is not a firm boundary and refers to the unlimited outer system. Inherit all assumptions from Example 4. Let $s_m$ and $d$ denote neighboring intersections' states (excluding $s_c$), and the traffic demand flowing to the inner system (the blue polygon) respectively. Then, the factored MOMDP can be formed as in Fig 1.12(b) with three parts: the controlled $s_c$, the observed and modeled $s_m$, and the observed only outer part $s_o$.*

*Here, the action and reward still only relate to the controlled part $s_c$, but not the inner system. And the $s_m$ as a subset of the outer system identified in Example 4 does not contribute to the value function directly. It is also worth noting that, the link between $s_c$ and $s_o$ can be omitted if the $s_m$ region completely encircles the $s_c$ region in Fig 1.12(a), that is, $s_c$ does not interact with $s_o$ directly.*

Example 4 can be seen as an extreme case of Example 5, in which the system boundary (the blue boundary) shrinks to the red and $S_m = \emptyset$.

The above single-intersection scenario can be extended to regional control with a centralized agent (controller) while all settings, assumptions, and the DBN representation in Fig 1.12(b) still hold true. For example, the red boundary in Fig 1.12(a) expanding to the blue give us a four-intersection centralized control model. Further, following the factorization idea introduced in Example 2, we can also extend Example 5 to a decentralized form which is analyzed in Section IV.B.

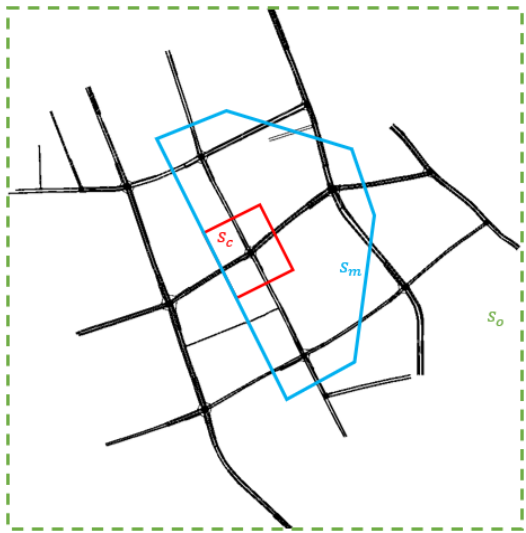
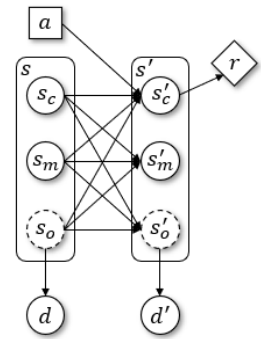

(a) An intersection within a traffic network        (b) The corresponding factored MOMDP representation

Figure 1.12. Formulate generic TSC problems as 3-stage factored MOMDPs

## IV. MODERN TSCs UNDER THE UNIFIED VIEW: SOLUTION MODELING

### A. Centralized Approximated Solutions to the MOMDP

Using the MOMDP representation introduced in Section III.A.4, we can unify the view of two main branches in modern optimized ATSCs: rolling-horizon optimization and model-free reinforcement learning. We consider an extended scenario of Example 4 and 5, in which we control a region. And we start from a centralized paradigm for clarity followed by a decentralized perspective and approach.

#### 1) Model-Free Reinforcement Learning

The theorem of reinforcement learning is developed upon the MDP. Hence, it is natural to describe RL methods using MDP languages. There are generally two types of approaches in MF-RL: value-based and policy-based methods. Value-based methods evaluate and estimate the expected return of each state-action pair, which is called the Q-value, and derive the optimal policy from Q-values. Similar to (7) and (8), we define policy $\pi$'s state-action value function and its Bellman equation as:

$$Q^\pi(s,a) = \mathbb{E}\left[\sum_{t=0}^{\infty} \gamma^t R(s_t, a_t)\right] = \mathbb{E}_{a \sim \pi}[R(s,a)] + \gamma \sum_{s' \in \mathcal{S}} P(s'|s,a) \sum_{a' \in \mathcal{A}} \pi(a'|s') Q^\pi(s', a'). \tag{12}$$

Reinforcement learning methods are designed to optimize a policy in the case that we don't have prior knowledge on the system model, that is, we don't know the system transition distribution $P$ and the reward distribution $R$. Value-based RL methods try to estimate Q-values iteratively from agent-environment interactions, which imply those unknown distributions. The classic Q-learning adopts temporal difference (TD) backup, stores Q-values in the Q-table explicitly, and forms a simple value iteration update:

$$Q'(s,a) \leftarrow Q(s,a) + \alpha \cdot \mathcal{L}_{TD} = Q(s,a) + \alpha \cdot \left(r + \gamma \cdot \max_{a}{}' Q(s',a') - Q(s,a)\right), \tag{13}$$

where $Q'$ is the updated new value; $\alpha$ is the learning rate; $r$ is the immediate reward from each sample; $\mathcal{L}_{TD}$ is the TD-error. DQN as its extension replaces the tabular representation with DNNs while keeping the TD-update framework unchanged. DQN variations can be found in (Hessel et al. 2018; Horgan et al. 2018).

Different from value-based methods, policy-based methods search in the policy space directly and easily support continuous action spaces, which are more challenging for value-based methods. The classic policy gradient (PG) tries to optimize the expected total return by observing experience trajectories $\tau = \{s_0, a_0, \dots, s_T, a_T\}$ directly:

$$\max_\theta J = v^{\pi_\theta}(s_0) = \sum_\tau R(\tau) \cdot p_\theta(\tau) = \mathbb{E}_{\tau \sim p_\theta(\tau)}[R(\tau)], \tag{14}$$

$$p_\theta(\tau) = p(s_0) \cdot \prod_{t=0}^{T} \pi_\theta(a_t|s_t) \cdot p(s_{t+1}|s_t, a_t), \tag{15}$$

where $\pi_\theta(a|s)$ is a parameterized behavior policy; $p_\theta(\tau)$ is the trajectory distribution under policy $\pi_\theta$; $R(\tau)$ is the total reward distribution over trajectory. Similar to the Q-learning setting, we still don't hold knowledge on distributions of transition $p$ and reward $r$ either. What's different is that PGs optimize the policy parameter $\theta$ directly through gradient ascent (the REINFORCE algorithm (Williams 1992)):

$$\begin{aligned}
\nabla_\theta V_\theta(s) &= \nabla_\theta \sum_\tau R(\tau) p_\theta(\tau) \\
&= \sum_\tau R(\tau) \nabla_\theta p_\theta(\tau) \\
&= \sum_\tau R(\tau) p_\theta(\tau) \frac{\nabla_\theta p_\theta(\tau)}{p_\theta(\tau)} \\
&= \sum_\tau R(\tau) p_\theta(\tau) \nabla_\theta \log p_\theta(\tau) \\
&= \mathbb{E}[R(\tau) \nabla_\theta \log p_\theta(\tau)].
\end{aligned} \tag{16}$$

By substituting $p_\theta(\tau)$ with (15), we can further simplify the log trajectory probability since the dynamic term $p(s_{t+1}|s_t, a_t)$ is independent from parameter $\theta$:

$$\nabla_\theta V_\theta(s) = \mathbb{E}\left[R(\tau)\nabla_\theta \log\left(\prod_{t=0}^{T} \pi_\theta(a_t|s_t) \cdot p(s_{t+1}|s_t, a_t)\right)\right]$$

$$= \mathbb{E}\left[R(\tau)\nabla_\theta \sum_{t=1}^{T} (\log p(s_{t+1}|s_t, a_t) + \log \pi_\theta(a_t|s_t))\right]$$

$$= \mathbb{E}\left[R(\tau)\sum_{t=1}^{T} \nabla_\theta \log \pi_\theta(a_t|s_t)\right]. \quad (17)$$

By doing Monte Carlo sampling, PGs approximate the optimum given by the expectation through averaging the experience:

$$\nabla_\theta V_\theta(s) \approx \frac{1}{N}\sum_{n=1}^{N} R(\tau^n) \cdot \left(\sum_{t=1}^{T} \nabla_\theta \log \pi_\theta(a_t^n|s_t^n)\right). \quad (18)$$

Other representative PG methods are advantage actor-critic (A2C) and extensions (Mnih et al. 2016), trust region policy optimization (TRPO) (Schulman et al. 2015), and proximal policy optimization (PPO) (Schulman et al. 2017). Deep deterministic policy gradients (DDPG) and its extensions combine value-based and policy-based ideas together and show state of the art performance on classic control tasks (Lillicrap et al. 2016; Barth-Maron et al. 2018).

*2) MF-RLs as the Solution*

For MF-RLs, either value-based or policy-based methods are maximizing the *expected discounted total reward* $\sum_t \gamma^t r_t$ of the controlled system $s_c$. We are further interested in how approaches delimit the system boundary, that is, how they define their *state spaces*:

1. Most of the literature defines the state space as the traffic status of the controlled region $s_c$ or its observation, for example, the joint of multi-intersection local state spaces. Without considering the influence of exogenous events, the system transition distribution described by $P(s'_c|s_c, a)$ can hardly meet either the Markov or the stationary assumption of MDPs. In this case, an MF-RL solution could face convergence, generalization, and distribution shifting issues at the same time.

2. Some approaches consider the external influence and include traffic demand into the state space $\mathbf{s} = [s_c, d]^\top$. This is the simplest solution to curbing exogenous events' impact. However, since the $s_o$ is not and can not be modeled, only observing $d$ still gives us a non-stationary state space. Because RL methods depend on the stationarity of all transition distributions, proper estimation of the demand provides more information to the decision-maker, but still leaves potential issues on the performance.

3. Approaches including neighbors' states into the agent's state space take the strategy of boundary expansion. They build the state space as $\mathbf{s} = [s_c, s_m]^\top$. As we discussed earlier, expanding the boundary can help mitigate the impact of exogenous events on the boundary of the new system. But, if the system boundary is not large enough so that exogenous events can be treated as white noises, this type of approach lacks the capability of handling the unexpected exogenous event.

As a conclusion, combining the optimal policy class analysis we made in Section III.A.3, MOMDP part, an ideal MF-RL solution to ATSC should optimize a belief-state-dependent policy. That implies a state space as $\mathbf{s} = [s_c, s_m, b_d]^\top$, where $b_d$ is the belief over the demand. Now, the implied transition distribution meets the MOMDP representation and leads to a potential optimal policy. A question left here is how to estimate $b_d$, or, the transition of $P(d'|d)$. We elaborate on these discussions in the proposed methodology part, see Section V.A. Also, how to handle the scenario that neighboring intersections are controlled by other agents is another challenge, in which case there will be an extra action term influencing the dynamic of $s_m$. This mostly happens in a decentralized control paradigm and will be covered in Section IV.B. The MARLIN ATSC system (Eltantawy and Abdulhai 2012) attempt to address this issue.

*3) Rolling Optimization and Model Predictive Control*

Rolling optimization is a combination of various ideas to solve optimization problems, including multiple areas using different languages. For clarity, we unify their representations and describe rolling optimization with the MDP terminology.

**Rolling-Horizon Optimization: Finding the Optimum**

RHOs operate with a *given system transition model* $P(s'|s, a, e) \in \mathcal{P}(\mathcal{S})$, in which $e$ representing the exogenous event is optional. Unlike RLs, RHOs give an opportunity to model the external influence explicitly. The transition model can be either probabilistic or deterministic as in MPC. It is also not limited to explicit dynamic models identified from real systems and can be approximated by ANNs and learned from data.

RHOs work with a fixed prediction horizon $T_p$ which is an integer multiple of the length of a decision epoch (time step). At each decision point $t$, the optimizer receives the newest system state $s_t$ and a predicted exogenous event vector till the end of the prediction horizon $\bar{\mathbf{e}}_t$, rolls out the system following the system model $P(s'|s,a,e)$ setting actions along the trajectory as unknown variables $\bar{\mathbf{a}}_t$, and produces the predicted system state vector $\bar{\mathbf{s}}_t$:

$$\begin{aligned}
\bar{\mathbf{e}}_t &= [\bar{e}_t, \bar{e}_{t+1} \dots, \bar{e}_{t+T_p-1}]^\mathsf{T}, \\
\bar{\mathbf{a}}_t &= [\bar{a}_t, \bar{a}_{t+1} \dots, \bar{a}_{t+T_p-1}]^\mathsf{T}, \\
\bar{\mathbf{s}}_t &= [\bar{s}_{t+1}, \bar{s}_{t+2} \dots, \bar{s}_{t+T_p}]^\mathsf{T}.
\end{aligned} \tag{19}$$

Also, RHOs require the prior knowledge on the reward function $R(\bar{\mathbf{s}}_t, \bar{\mathbf{a}}_t)$, which is an objective function over the prediction horizon. Formally, we can define the optimization problem as follow:

$$\max_{\bar{\mathbf{a}}_t} \mathbb{E}[R(\bar{\mathbf{s}}_t, \bar{\mathbf{a}}_t)],$$
$$s.t. \begin{cases} \bar{s}_t = s_t \\ P(s'|s,a,e) \in \mathcal{P}(\mathcal{S}) \\ \bar{s}_t \in \mathcal{S}, \bar{a}_t \in \mathcal{A} \\ other\ action\ constraints. \end{cases} \tag{20}$$

By solving (20), we obtain the optimal action trajectory $\mathbf{a}^* = [a_t^*, \dots, a_{t+T_p-1}^*]^\mathsf{T}$ for the next prediction horizon. We pick the first action $a_t^*$ in the vector, execute it, drive the system one step forward, and repeat the above procedure till the end of the episode. Fig 1.13 shows a systematic diagram of RHO's online optimization framework, where $\mathcal{J}_t = \mathbb{E}[R(\bar{\mathbf{s}}_t, \bar{\mathbf{a}}_t)]$.

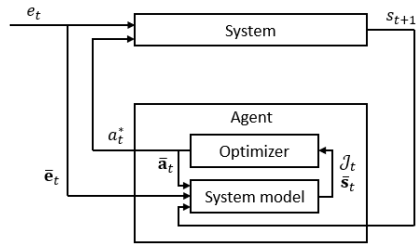

Figure 1.13. The online optimization structure of general rolling-horizon optimization methods

**A Second Glance at the Control: Optimal Control and MPCs**

We have covered the foundation of classical control theory in Section III.A.1, showing that the major objective is to guarantee the system stability through the control law design, while maintaining only a degree of optimality, such as the speed of response. As the controlled system and control target becoming more complex, e.g., a nonlinear system with constraints and a loss function considering energy consumption, the optimal control was introduced (Luenberger 1979, 393–435). It shares the same foundation we stated before, but tilts the major focus on finding the optimality. Hence, it is naturally highly related to the optimization research. And many classic books discuss their connection, from either fields' perspective (Bertsekas 2019).

In optimal control theory, MPCs, which dealt with constrained problems with uncertainty well, is a good example. It builds upon the idea of rolling (receding) optimization (Xi and Li 2019, 1–10), and extends it to close the gap between control performance criteria (stability, robustness, etc.) and RHOs. MPCs share the same optimization methodology with RHOs, hence also apply to (19), (20), and Fig 1.13.

Still, in this article, we do not focus on the stability proof but rather focus our interest on the analysis of the system structure. Following the taxonomy introduced in Section III.A.1, we observe the policy's dependency. Fig 1.13 shows that, the agent makes decisions based on both the original system input (exogenous events) $\bar{\mathbf{e}}^t$ and the state $s^t$, which implies a structure of *feedforward-feedback combination*. Fig 1.14 translates the workflow into a block diagram and gives us more insights on the structure.

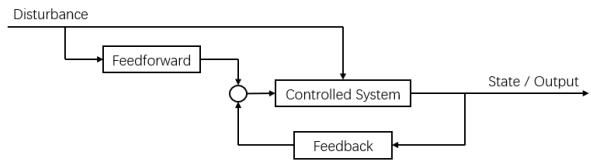

Figure 1.14. The block diagram of MPCs. The target value branch is omitted since there is no set point in this optimal control setting.

Compared with the pure feedback MDPs, MPCs handle exogenous disturbances directly with the feedforward signal. But it actually meets with MDPs from the perspective of problem modeling. The feedforward signal commonly relies on another prediction model. For example, in the TSCs scenario, imagine that the traffic management center feeds the peripheral network's traffic status to a controlled intersection, which is given by a separate model (partially) describing the outer system (or the whole system). On another hand, from the perspective of MDPs, if needed, we can re-delimit the system boundary and model the significant part of the outer system together. Hence, they are just two approaches to a common question. Surely, the separate predictor can likely introduce an estimation error to the whole control system, especially in a decentralization framework. But a similar issue is also encountered by RL solutions since the MDP implied by the agent design commonly varies from the true one.

*4) RHOs as the Solution*

From the objective perspective, RHOs are maximizing the expected total reward *give a horizon* $\sum_{t=0}^{T} r_t$. Compared with RLs' target, this horizon-limited objective seems more myopic, but not necessarily, since after a long enough horizon, the discounting term in RLs' target converges to 0. And without the discount factor, RHOs' target grants equal weights to rewards in the near future.

When a system boundary is identified, RHOs model the system dynamic considering exogenous events at the mean time $P(s'_c|s_c, a, d)$. It is different from the second type in Section IV.A.2: RHOs commonly have an $d$ prediction model that is independent from the system dynamic model. Such a demand prediction model can vary in realization by much, which could inherently consider the dynamic of $s_o$, for example, a traffic prediction model corresponding to the larger area of traffic networks. Feeding the prediction of future exogenous events allows RHOs to handle non-stationary $d$ by optimizing through the trajectory of predicted future $\bar{\mathbf{d}}$ completely. Hence, even RHOs share a common implied MOMDP model with RL methods, they utilize an online optimization approach with the advantage of handling the non-stationarity. However, if the prediction model in inaccurate, the control will also be inaccurate. Hence, the presence of a good prediction model in RHO maybe advantageous to no model as in MFRL approaches, but no-model can be better than an inaccurate model. On the other hand, this ability to seek optimal control in RHO comes at the expense of additional computational complexity of both the prediction model and the online optimization.

## B. Decentralized Approximated Solutions to the Factored MOMDP

In the case that the system and the modeled space $\mathcal{S}_c, \mathcal{S}_m$ are intractably large, we can further factorize the system using the FMDP idea. Decentralized ATSC methods commonly decompose a controlled region into intersections and assign distributed agents to each of them.

Here, we have to make a disambiguation, that a decentralized ATSC method may not be a formal solution to a decentralized (PO)MDP problem (Dec-POMDP). In a Dec-POMDP, agents observe and take actions locally, but the system state transition and reward distribution are jointly evolved. In which case, a Dec-POMDP solution has to be an *observation-history-dependent joint deterministic policy*. Solving such a problem explicitly is unrealistic since Dec-POMDPs are proved as NEXP-complete problems (Bernstein et al. 2002). Hence, most of Dec-POMDP solutions are seeking an approximated optimal policy class by simplifying the model and restricting the observation and transition dependencies. The relaxation on modeling leads the optimized policy to *observation-history-dependent independent probabilistic policies* since local agents no-longer model the global dynamic and lose the guarantee of optimal deterministic policy given by Markov and stationary settings.

We consider decentralized ATSC methods as simpler approximated solutions to Dec-POMDPs, which turn them into independent MOMDPs with global reward objectives and derive independent deterministic policies. Specifically, we use the two-intersection control extension of Example 4, which presumes $\mathcal{S}_m = \emptyset$ for simplicity. And factorize the $s_1$ into two intersection-related components. We derive an example as follow:

**Example 6** *We control the two intersections in solid rectangles in Fig 1.15(a). The joint of the two local state space forms the $s_c$ in Example 4, $s_c = (s_{c,1}, s_{c,2})$. And $s_o$ still denotes the outer system. Inherit all assumptions from Example 4. The factored MOMDP can be formulated as in Fig 1.15(b).*

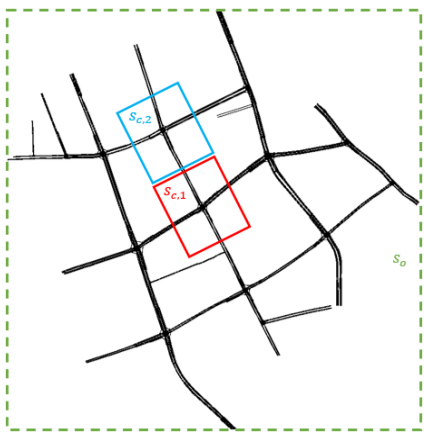
(a) Two intersections within a traffic network

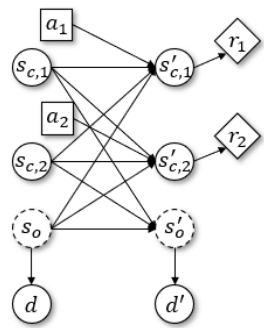
(b) The corresponding factored MOMDP representation

Figure 1.15. Formulate a two-intersection decentralized TSC problem as a factored MOMDP. The modeled-only region is omitted for simplicity

*1) Naive Decentralized Solutions*

We consider that controllers make decisions based on the same type of information in both training (learning or designing) and deploying. In this case, the decentralization relaxes the problem and sets up respective "centralized" controllers, each responsible for one intersection. These individual controllers generally fall into the discussion in Section IV.A, with minor differences. Without loss of generality, let's focus on a single agent of intersection 1. Here, we cannot simply merge $s_{c,2}$ into $s_o$ and treat them as a stationary peripheral network as before, since $a_2$ corresponds to another (non-stationary) policy and introduces more uncertainty to the system. In order to solve this new problem properly, we need to hold a belief to not only the exogenous events $d$, but also the neighbors' policy $\pi_2$. The discussion of our proposed solution is in Section V.1.

*2) CTDE Solutions*

Typically, RLs have the potential giving controllers access to more information during training than after deployment, which implies the centralized training and decentralized execution (CTDE) framework. In ATSC applications, centralized training has two purposes: training homogeneous agents for simplicity, and supervising the learning with global information for effectiveness.

The first target is commonly achieved by adopting graph attention networks (GATs) (Wei, Xu, et al. 2019). In GAT based ATSCs, each agent controls a single intersection while considering *n*-hop neighbors' observation (*n* depends on the number of GAT layers). Different from purely neighboring information sharing strategies, a GAT network with the attention mechanism can be adopted to sub-networks with various topology, which enables us to train a homogeneous policy/value approximation across all agents. Focusing on each single agent, the state space design majorly shares the same discussion in Section IV.B.1. But the centralized training brings differences. Different from the second type of CTDE method (covered later), agents in GAT based methods access the same information in both training and deploying, owing to the learned homogeneous mapping. The homogeneous mapping is a double-edged sword. This sharing means that the policy is trained on the experience with more plentiful patterns of dynamics and hence is expected to handle more testing scenarios. To the contrary, the abundant dynamics also involve exogenous influences from different sub-networks, which will hinder the learning and exacerbate the fore-mentioned non-stationary issue if the individual agent is not well-designed. Under this discussion, the promising policy sharing technique can be dangerous since it takes average over the experience of unexpected dynamics across the network.

The second purpose can be achieved by other RL techniques — value decomposition and global critiquing (under the actor-critic framework). They have greater differences from previous methods. During the training, agents' heterogeneous policies are supervised by global reward together with the fed global observation or even state for better coordination (reaching the global optimum). After deployment, the global supervising is abandoned. And agents execute learned policies locally. The agent design still shares the same discussion as in Section IV.B.1. While the centralized training procedure now assigns agents different targets. It is an effective technique for improving coordination capability (Zhang, Yang, and Zha 2020; Xu et al. 2021), but still suffers from issues like distribution shifting and non-stationarity. Meanwhile, a CTDE framework may meet more problems in real-world implementation, see Section V.A. In conclusion, CTDE has the potential to bring us a level of coordination, with a precondition of careful agent design.

*C. Extension: Classic TSCs Under the View of Control Systems*

At this point in the review, one question may arise: besides those modern optimized control methods, how can more classic TSC methods be fitted into and critiqued under the MDP framework? Unfortunately, not all those methods can be described since MDPs provide a framework for optimization problems. For example, SCOOT's online optimization based on recent collected data, can be fitted into the unified view. On the other hand, fixed-time controllers and actuated controllers follow a pre-defined heuristic strategy (control law) and commonly do not form or solve an optimization problem (the Webster method which tries to minimize the average intersection delay through a set of empirical equations can be seen as an exception). It is hard to evaluate them together with ATSCs. However, either of them can find a place under the view of control systems, which has been covered in Section III.A.1. In this section, we try to discuss and evaluate historical TSC methods (Section II.B) under our greater unified view.

*1) Fixed-time Control*

Fixed-time control has a structure of open-loop control. Their phase structure and sequence are commonly manually assigned based on the knowledge of the controlled system, such as the network layout and the historical demand. Like other open-loop control systems, keeping control unchanged make the system unable react to any unexpected real-time disturbances. This type of methods fall short from the optimal control problem and cannot be fitted into our MOMDP framework.

*2) Actuated Control*

Actuated control introduces the real-time adjustment functionality and can hence be considered as a closed-loop structure. The feedback control signal, i.e., extending or shortening a phase duration, grants it a level of ability to handle the fluctuation. But the feedback structure itself is not a sufficient condition for reaching either stability or optimality. Such a type of heuristic feedback control law without any proof only has a limited efficiency for encountering disturbances in the field. This type of method does not fit into the MOMDP framework as well.

*3) Adaptive Control: 1-st and 2-nd Generations*

TSCs became more mature after entering the adaptive control era. ATSCs introduce the objective of optimality under real-time disturbance and pursue higher control time granularity (how frequently the control adapts to changes). We've covered the most representative methods in Section IV.A and review their predecessors in this part.

1G ATSCs prepare multiple timing plans and select the most proper one according to the time of the day and the sensed traffic status. Similar to actuated control, they have the basic characteristic of a feedback structure. However, at this stage, controllers are limited by the computational and sensing power, i.e., both the control time granularity and control law are too coarse to react to high-frequency disturbance. As one classic feedback control method teaches us — bang-bang control — limited controller may bring optimality under the specific setting, may also bring poor performance and even system oscillation. Hence, either stability or optimality can hardly be met by 1G ATSCs.

The development of sensing, communication, and computation technologies brings 2G ATSCs more capability in encountering real-time disturbances: doing more fine-grained state-feedback, accessing more abundant network-level information, and optimizing online. From the perspective of a feedback control system, they can do better than 1G ATSCs. Also, 2G ATSC introduced a breakthrough hierarchical control paradigm. Such hierarchical paradigm is yet to emerge in modern optimized ATSCs. One major reason is that modern algorithms commonly operate in a much more smaller time granularity than those 2G ATSCs. As we stated before, given a time granularity, we can ignore an impact far from present in either time or space. Hence, a carefully designed decentralized method can also approximate the MOMDP model well (Section IV.B). While, in 2G ATSCs, the courser time granularity requires more efforts in reaching coordination.

Certainly, modern ATSCs have a higher opportunity in dealing with high-frequency disturbances due to the finer control law and finer objectives — optimizing at the level of seconds over the future. The way 2G ATSCs deals with large-scale coordination is still a fantastic idea that should inspire modern ATSC. Accordingly, a potential research direction is proposed in Section V.C.

## V. POTENTIAL DIRECTIONS: SUGGESTED BY THE UNIFIED VIEW

It is crucial for traffic system developers to design ATSCs with enough capability on both reliability and efficiency. Reliable means the controller's capability of generalization to different traffic volumes, which is naturally met by RHO methods. While efficient refers to the computational complexity in online deployment, which is a major advantage of MF-RLs. To better conduct cutting-edge research on enhancing both characteristics, it is important to resolve the critical questions the unified view pointed out.

### A. Solving Partial Observability and Non-stationarity Issues with Exogeneous Disturbance Modeling

According to the analysis in Section IV.B, while reducing a Dec-POMDP into independent MOMDP problems, it is crucial to identify the system boarder and handle the exogenous event carefully. Plentiful recent research investigates how to coordinate intersections better with deep reinforcement learning techniques. And the most popular direction might be the CTDE framework (centralized training, decentralized executing, see Section II.B.2, MF-RL part). By combining graphical neural networks (GNNs), end-to-end CTDE-based DRL methods are expected to reach autonomous agent-correlation (considering assigning different correlation factors to neighbors during coordination) recognizing, global information sharing, and fully decentralized deployment simultaneously. However, regardless of whether these wishes become a reality, the assumptions CTDE relies on could not be always fulfilled in-field. In ATSC, traffic light agents are commonly expected to adapt to varying traffic demand. That requires a solution either having an online optimization capability that is based on the newest detection, or at least maintaining online strategy-updating capability. While CTDE assumes deploying fixed distributed policies and requires global knowledge for policy-updating. Since MF-RLs are not guaranteed to be reliable in responding to unforeseen experiences, abandoning the updating capability might be dangerous.

It suggests that MF-RL ATSCs should preserve the decentralized paradigm to reach the scalability.

Section IV.B provides a guideline for designing decentralized multi-agent MF-RL agents. Commonly, literature models each agent with its corresponding intersection only, in which the incoming traffic demand is the exogenous disturbance. As analyzed before, to mitigate the non-stationarity that exogenous events bring, a trade-off between model complexity and efficiency should be made. A simple but effective approach is to model the system-wise interaction as a MOMDP as introduced in Example 4. We treat the local intersection state as the fully observable component, and the incoming demand as the partial observation of the outer system. To find its optimal history-dependent deterministic policy, the state space of MF-RL should include the current local state and the history of incoming traffic demand readings.

The demand history can be replaced with some latent embeddings for simplicity. To explain the feasibility of such a simplification, we take the notation in Example 4. From the perspective of modeling a solution, we do not model the outer system $s_2$ and are interested in estimating the transition distribution $P(d'|d)$ from the experience. We can degenerate the joint, stationary dynamic of $s_2$ and $d$ to a non-stationary process $\{d_t\}$ solely. In each time step, we take an observation $o_t$ from $\{d_t\}$. This generally gives us a hidden Markov model (HMM), as shown in Fig 1.16. And our object is to do the Bayesian inference to the hidden (non-stationary) process from the observation history. This objective can be easily achieved by classic models like Kalman filtering and RNNs.

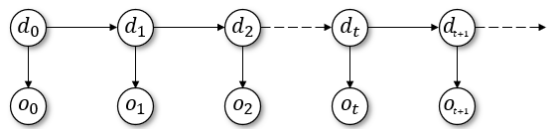

Figure 1.16. A hidden Markov model, where the hidden process is non-stationary.

The above method neglects the non-stationarity from neighbors' policies and tries to average them. To resolve this limitation, MARLIN (El-Tantawy, Abdulhai, and Abdelgawad 2013) models neighboring intersections' dynamic, introduces non-stationary converging policies (NSCP) to estimate the influence from neighbors' actions, and hence captures exogenous disturbances. Naturally, it is promising to extend the tabular MARLIN with DRL techniques. Despite DNNs can handle the joint spaces easier than tables, it is still an interesting research direction to approximate neighbors' time-varying polices smoothly by DNNs.

*B. Alleviating Distributional Shift Issues with Model-Based Online Optimization*

According to the comparison in Section IV.A, online optimizations (RHOs) with extra outer-system dynamic prediction models hold advantages in handling exogenous disturbances. It suggests improving the MF-RL policy generalizability by adding the RHOs' online optimization scheme.

The root cause behind the reliability issue of the MF-RL learned policy is that the fixed policy cannot describe the full decision space owing to the limited visitable area during training restricted by the experience. The probabilistic distribution representing the experienced system dynamics will change from training to testing according to the corresponding outer system's dynamics (e.g., neighboring intersection policies and the peripheral network-generated traffic demand). The learned policy can only respond appropriately to those frequently visited states, and may likely fail to make correct decisions within undertrained parts of the search space. This distributional shift issue limits RL policies' generalizability over different demand patterns, even if they fall within the model implied by the MF-RL agent design. RHO type of online planning utilizing value approximations given by MF-RLs has the potential in alleviating such an issue. Given a model describing the system dynamics, further planning can smooth out poor value estimations in the state space, help the controller in making better decisions, and hence improve the reliability of the control system.

This is a natural idea, but not easy to achieve in reality. Improper design or an extremely poor planning model may lose efficiency and reliability at the same time. A close idea can be found in the field of model-based reinforcement learning (MB-RL). MB-RLs share the same assumption as MF-RLs and try to *learn the system dynamic model first* from interacting with the environment rather than learning the value function or policy directly. Some MB-RLs utilize the learned model to generate more experiences (Sutton 1991); while others optimize the future and do planning on it (Amos et al. 2018). Jaggi et al. (2021) introduces a model-based Monte Carlo tree search (MCTS) approach and indicates that the agent can generalize to unexperienced demand profiles better with a model. But learning a good-enough model from interactions is again a challenge since that inadequate or even biased experiences (interaction data) induce unreliable models which cannot lead the planning to the expected optimum.

*C. Improving Large-Scale Network Coordination with Hierarchical Paradigm*

The former two parts maximize the coordination capability of the decentralized paradigm by imposing careful model designing. But the relaxation from Dec-POMDP to independent MOMDPs means that the global optimal can't be reached. On the other hand, endlessly expanding the system boundary is expensive and might be helpless. This is affected by both the network scale (Section III.A.4) and the time granularity. Hence, beyond the scope of small scale regional control that controls a small set of close-distanced intersections, we need some different strategies for coordinating large-scale traffic networks.

A hierarchical paradigm is a potential tool. As defined in Section II.A.3, hierarchical control means multi-level heterogeneous controllers working on different geometric scope and time granularity, and interact with each other. Although (Kouvelas, Triantafyllos, and Geroliminis 2018) form heterogeneous controllers responsible for different intersections and claim the approach as "hierarchical", we still classify them into the decentralized category. Different from these, we can have inter-region controllers as high-level agents in the hierarchical paradigm and design regional ATSCs as low-level agents. The high-level controller operates in a larger time scale, considers large-space traffic dynamics, and poses control schemes or action constraints to the lower levels. Major benefits of doing this include decoupling the duty of large-timescale, large-space coordination from ATSC agents and alleviating the complex modeling requirement of ATSC.

Besides, identifying regions' boundaries is also a challenging problem. It is intuitive to handle it manually according to the history observation, but the rapidly varying traffic dynamics may create different critical zones in different periods of a day. This issue implies the requirement on dynamic region boundary identification. More generically, we can assign intersection-level agents different roles when they are on/off the boundary of a region (Xu et al. 2020). We might be able to identify their roles with learning-based methods, such as the role-oriented MARL (ROMA). By interacting with the simulator, ROMA may learn to classify agents by their features dynamically.

# APPENDICES

## I. TSC STATE REPRESENTATIONS

Various information can be used to form ATCS's state space. Commonly used state components include:

- *Vehicle speed* $v_i$ of vehicle $i$ on incoming edges.
- *Vehicle position* $z_i$ of vehicle $i$ on incoming edges. Commonly, it is the vehicle's relative distance to the stop bar and ignores the lateral displacement. Together with the vehicle speed, these two state components can be measure by either surveillance cameras or V2X communications.
- *Traffic density* $\rho_i$ of section (or incoming lane, edge) $i$.
- *Space mean speed* $u_i$ of section (or incoming lane, edge) $i$.
- *Traffic volume (flow)* $f_i$ of section (or incoming lane, edge) $i$.
- *Vehicle count* $c_i$ of incoming lane (or edge) $i$.
- *Queue length* $q_i$ of incoming lane (or edge) $i$. A vehicle is counted as waiting in the queue when its speed is below a threshold $v_i \leq v_{th}$ ($v_{th} = 2m/s$ in this research).
- *Phase* of the traffic light. It can include both the current phase an the upcoming phases.
- *Progress of the phase*. It indicates the ratio of current phase duration and minimum/maximum allows phase duration. It also provides a clue of available actions to the agent.
- *Traffic demand* from an incoming edge. It can be measured by upstream loop detectors, or be estimated from neighboring intersections' dynamics.
- *Discrete traffic state encoding (DTSE)* of incoming edges. It discretizes the space into cells, as shown in Fig. 17, where we record the number of vehicles presenting in each cell and their average speed. This representation also requires either surveillance cameras or V2X communications.
- *Temporal discrete traffic state encoding (TDTSE)* of incoming edges. This representation records the presents of vehicles on loop detector groups as in Fig. 18. It does not rely on pioneer sensing technologies, but provide lesser traffic status. Hence, TDTSE expands on the time axis filling up missing information. Both DTSE and TDTSE are developed for adapting CNNs.

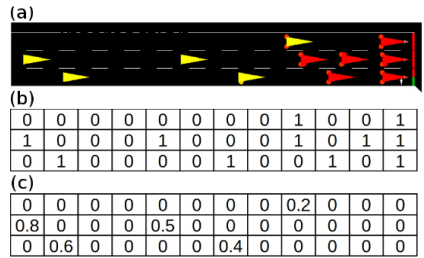

Figure 1.17. An illustration of the DTSE representation. (a) An approach of an intersection. (b) The occupancy matrix. (c) The normalized vehicle velocity matrix. Adapted from (Genders and Razavi 2016).

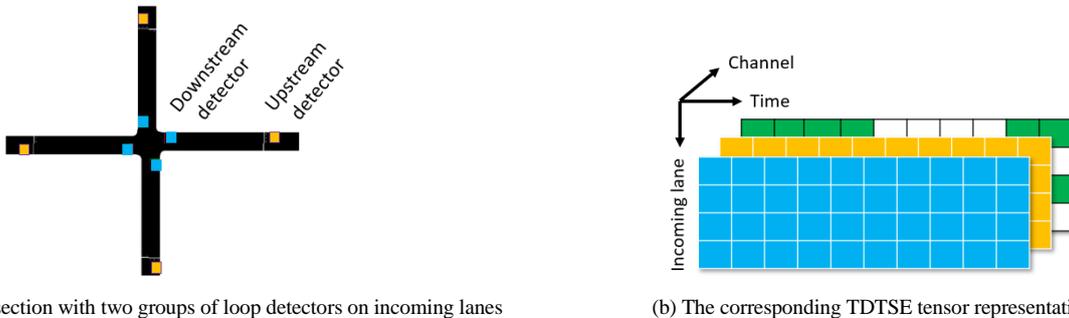

(a) A 4-arm intersection with two groups of loop detectors on incoming lanes    (b) The corresponding TDTSE tensor representation

Figure 1.18. An example of the TDTSE representation. Matrices in the TDTSE tensor represent channels, where first two channels correspond to the two groups of detectors and the last channel indicates the phase history. Cells in detector channels contain binary numbers indicating the presence of vehicles.

## II. TSC Criteria

### A. Intersection-Level Criteria

First, we introduce intersection-specific metrics. To be noticed, the delay introduced below is different from the traditional definition in the traffic field, which is the difference between the actual travel time and the ideal travel time (corresponding to the free-flow speed) (Dowling 2007). Most TSC literature defines the "delay" as "stop time" which indicates the time of vehicles driving below a threshold speed. When we calculate it every time step, the stop time is equal to the number of stopped vehicles in quantity. Hence, in this article, we let the queue length, stop time, and delay share the same definition, while rename the formal "delay" as the "actual delay".

- *Queue length (stop time, delay)*: the total queue length of an intersection:
$$Q^t = \Sigma_i q_i^t.$$

- *Actual delay*: the actual time loss compared to the ideal travel time:
$$AD^t = \tau \cdot \Sigma_i \frac{v^* - \bar{v}_i}{v^*},$$
where $\tau$ is the length of a time step, $v^*$ is the free-flow speed, and $\bar{v}_i$ is the average speed of vehicle $i$ in the last time step.

- *Number of stops*: the total number of stops vehicles take in one time step.
$$NS^t = \Sigma_i S_i^t,$$
where $S_i^t$ is the number of stops vehicle $i$ took last time step. A stop behavior is identified when a vehicle decelerates below the threshold speed. This criterion reflects the quality of drivers' experience.

- *Cumulative delay* (El-Tantawy, Abdulhai, and Abdelgawad 2013): the cumulative total delay (stopping time) of all vehicles on the incoming links:
$$CD^t = \Sigma_i D_i^t,$$
where $D_i^t$ is the total delay vehicle $i$ wasted within the intersection's detection area. It increases over time once vehicle $i$ joins the detection area and is reset to 0 once it leaves the intersection (passes the stop line). Compared with the queue length, the cumulative delay increases rapidly when vehicles waiting before the stop line for a long period because it integrates total stop times over all vehicles. This metric tends to give more penalty to the decision holding vehicles for a long period.

- *Vehicle count*: the total number of vehicles on all incoming links:
$$VC^t = \Sigma_i c_i^t,$$

- *Cumulative travel time*: the cumulative total travel time of all vehicles on the incoming links:
$$CTT^t = \Sigma_i T_i^t,$$
where $T_i^t$ is the travel time vehicle $i$ spent within the intersection's detection area.

- *Difference in cumulative delay* (El-Tantawy 2012, 116–17): the difference of the cumulative delay in two consecutive seconds:
$$DCD^t = CD^t - CD^{t-1}.$$

We can also define a metric as a combination of traffic states. For example, a commonly used cost function in the traffic field is:
$$Cost^t = a \cdot Q^t + b \cdot NS^t,$$
where $a$ and $b$ are two coefficients.

### B. Network-Level Criteria

For the multi-intersection scenario, one simple approach is to sum a certain metric over intersections. Besides, criteria defined at the network level directly to evaluate global performance are listed below:

- *Average travel time*: the average travel time of all completed trips:
$$ATT = \frac{1}{N} \sum_{i=1}^{N} T_i^{\mathcal{G}}, \quad c_i \in \mathcal{C}^{\mathcal{G}}, \quad |\mathcal{C}^{\mathcal{G}}| = N,$$

where $\mathcal{C}^{\mathcal{G}}$ is the set of vehicles that completed their trips, $T_i^{\mathcal{G}}$ is the travel time of the completed trip.

- *Average delay*: the average delay of all completed trips. Notice, here the delay also refers to the stopping time.

$$AD = \frac{1}{N}\sum_{i=1}^{N} D_i^{\mathcal{G}}, \quad c_i \in \mathcal{C}^{\mathcal{G}}, \quad |\mathcal{C}^{\mathcal{G}}| = N,$$

where $T_i^{\mathcal{G}}$: the stopping time (delay) of the completed trip.

- *Throughput* counts the number of completed trips:

$$Throughput = |\mathcal{C}^{\mathcal{G}}|.$$

C. *Arterial-Level Criteria*

Finally, we introduce arterial-level indicators designed to evaluate adjacent intersections' signal progression quality.

- *Arrival type* (Manual 2000) is an indicator with 6 levels describing the progression of a link **qualitatively**. All 6 levels' definitions are listed in the last two columns of Table 1.1. The definition of arrival type is intuitive, but it is hard to measure quantitatively: researchers have to observe the traffic flow in the field and assign arrival type to each link based on what they perceived. Hence, researchers introduce the platoon ratio to **quantitatively** measure the signal progression, from sensors readings.

Table 1.1. Relation between the arrival type and the platoon ratio

| **Platoon Ratio** | **Arrival Type** | **Description of Flow** |
|---|---|---|
| 0.333 | 1 | Very poor progression |
| 0.667 | 2 | Unfavorable progression |
| 1.000 | 3 | Uncoordinated signals or random arrivals |
| 1.333 | 4 | Favorable progression |
| 1.667 | 5 | High favorable progression |
| 2.000 | 6 | Exceptional progression |

- *Platoon ratio* is the ratio between two quantities: the proportion of vehicles arriving during the green phase in a full cycle, and the proportion of the green light duration in the cycle time. We can also derive the platoon ratio from the proportion of two flow rates:

$$R_p = \frac{P}{g/c} = \frac{f_g}{f},$$

where $P$ is the ratio of vehicles arriving during the green phase, $g/c$ is the ratio of green light duration over the cycle time, and $f_g$ and $f$ are the flow rates during the green light period and the whole cycle respectively. The higher platoon ratio means more vehicles that come from the upstream intersection could pass by the downstream one without stopping, which indicates better coordination between the two intersections. The relation between the platoon ratio and the arrival type is given by Table 1.1.

- *Green band width (in second)* describes the period allowing platoons passing by a corridor without stopping. Fig 1.19 illustrates double-direction green bands of a three-intersection corridor on a time-space diagram. The green band is determined by cycle lengths, offsets, and the average platoon speed. In Fig 1.19, the slope of a green band represents the platoon speed. $b$ and $\bar{b}$ are northbound and southbound green band widths correspondingly. We consider the proportion of the bandwidth to the cycle length $b/C \times 100\%$ as the progression efficiency. The higher efficiency the better.

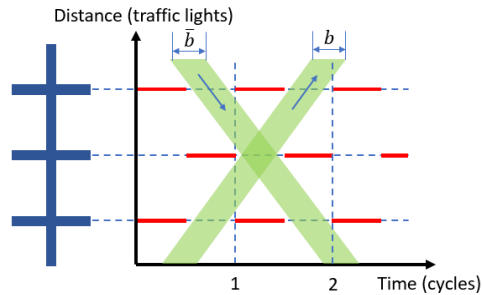

Figure 1.19. Green bands in a time-space diagram.